\documentclass[aps,prd,twocolumn,superscriptaddress]{revtex4-2}

\usepackage{makecell} 
\usepackage{microtype}
\usepackage{graphicx}
\usepackage{dcolumn}
\usepackage{bm}
\usepackage{lmodern}
\usepackage{caption} 
\usepackage{mathtools} 
\usepackage{tensor} 
\usepackage[hidelinks]{hyperref} 
\usepackage{amssymb}
\usepackage{mathtools}
\usepackage{mathrsfs}
\usepackage{amsmath} 
\usepackage{xcolor}
\usepackage{orcidlink} 
\usepackage{array} 
\usepackage{booktabs}
\usepackage{subcaption}
\usepackage{enumitem}
\usepackage{soul}
\usepackage{amsthm}
\usepackage{stmaryrd}

\usepackage{amsfonts} 

\newcommand{\Ree}{\mathfrak{R}}
\newcommand{\Imm}{\mathfrak{I}}

\usepackage[toc,page]{appendix}

\newtheorem{theorem}{Theorem}

\theoremstyle{remark}

\newcommand{\dd}{\mathrm{d}}

\newcommand{\s}{\mathfrak{s}}

\newcommand{\schdual}{\widetilde{\mathfrak{s}}}
\newcommand{\al}{\alpha_0}

\AddToHook{shipout/before}{%
  \pdfpageattr{/Rotate 0}%
  \ifnum\value{page}<\getpagerefnumber{tab:catalogue}\relax\else
    \ifnum\value{page}>\getpagerefnumber{tab:catalogue:end}\relax\else
      \pdfpageattr{/Rotate 90}%
    \fi
  \fi
}

\begin{document}

\title{Harmonic branches of Einstein--ModMax--scalar field gravity: exact solutions and regularity II}

\author{Leonel Bixano\,\orcidlink{0009-0009-5858-6466}}
    \email{Contact author: leonel.delacruz@cinvestav.mx}
\author{Tonatiuh Matos\,\orcidlink{0000-0002-0570-7246}}%
 \email{Contact author: tonatiuh.matos@cinvestav.mx}
\affiliation{Departamento de F\'{\i}sica, Centro de Investigaci\'on y de Estudios Avanzados del Instituto Politécnico Nacional, Av. Instituto Politécnico Nacional 2508, San Pedro Zacatenco, M\'exico 07360, CDMX.
}%

\date{\today}

\begin{abstract}
We integrate harmonic solution branches of Einstein--ModMax--Scalar field gravity, keeping the electromagnetic phase \(\tan{\Theta}=\mathcal{G}/\mathcal{F}\) non-constant. For each family we first solve the field equations $(a(\s),b(\s),c(\s))$ and reconstruct the physical functions $f,\omega,k,A_t,A_\varphi,\phi$. In the same section we then select the Schwarzschild or Coulomb harmonic function, impose the regular horizon conditions, identify the ModMax solution, and state its Maxwell limit. Two theorems distinguish a constant electromagnetic phase from an exact static charge rescaling and show how twist prevents constant diagonal rescalings of the electromagnetic potentials at fixed metric and scalar. The catalogue covers known RN and charged Taub--NUT solutions, single-charge configurations, explicit static dyons with scalar-dependent couplings, and a restricted stationary family with an active scalar. We distinguish new explicit ModMax extensions of known Maxwell profiles from the nonfrozen stationary integration. The axial LWP reconstruction recovers the known Melvin--swirling background and gives an exact ModMax Schwarzschild exterior within that background. At its horizon, smooth metric and electromagnetic limits are distinguished from the constitutive continuation at a null equatorial circle. Uncharged configurations and earlier Maxwell fields are traced back to the preceding harmonic catalogue.
\end{abstract}

\maketitle

\section{Introduction}
The Ernst formalism connects stationary, axisymmetric gravitational fields to a set of equations for potentials \cite{Ernst:1967by,Ernst:1967wx}. The harmonic potential approach decouples the integration of these equations from the later selection of the harmonic function \cite{Matos:1995my,Matos:2000za}. In Einstein--Maxwell--scalar field gravity, this framework yields an explicit link between the integration constants, the conserved energy in potential space, and the regularity of the reconstructed fields \cite{Bixano:2026xjg}. In this work we apply the same strategy to Einstein--ModMax--scalar field gravity, using the potentials and conventions of Ref. \cite{Bixano:2026ouq}, which is an extension of the original work \cite{Bixano:2026xum}.

ModMax is a nonlinear generalization of Maxwell electrodynamics that maintains electromagnetic duality and conformal symmetry \cite{Bandos:2020jsw}. When coupled to gravity, it allows for a rescaled Reissner--Nordstr\"om geometry \cite{Flores-Alfonso:2020euz,Amirabi:2020mzv}, charged Taub--NUT solutions featuring a nonconstant electromagnetic phase \cite{BallonBordo:2020jtw,Flores-Alfonso:2020nnd}, and various axisymmetric setups \cite{Barrientos:2022bzm,Barrientos:2024umq}. Bokuli\'c and Herdeiro derived nonextremal multi--black-hole configurations together with a balanced extremal limit \cite{Bokulic:2025usc}, and they also broadened solution-generating techniques to include dilatonic cases \cite{Bokulic:2025ucy}. These results must be identified explicitly before a harmonic representative can be called new.

Recent Maxwell constructions provide further comparisons for the axial representations \cite{DiPinto:2025yaa,Astorino:2026kuv}. Stationary ModMax solutions with a conformally coupled scalar are also known \cite{Ayon-Beato:2024vph}. Their scalar coupling differs from Eq. \eqref{eq:action} and is distinguished from the dilatonic families below.

Unless stated explicitly to the contrary, we use the conventional scalar sign choice, $\epsilon_0=+1$. The action is
\begin{equation}
     \begin{split}
     S={1\over16\pi}\int\dd^4x\sqrt{-g}\,
     \big[&-R+2\epsilon_0(\nabla\phi)^2\\
     &+e^{-2\al\phi}\mathcal L_{\rm MM}\big],
     \end{split}\label{eq:action}
\end{equation}
where
\begin{equation}
     \mathcal L_{\rm MM}=\mathcal F\cosh\gamma+
     \sqrt{\mathcal F^2+\mathcal G^2}\sinh\gamma,
     \label{eq:modmax}
\end{equation}
$\mathcal F=F_{\mu\nu}F^{\mu\nu}$, $\mathcal G=F_{\mu\nu}{}^\star F^{\mu\nu}$, and $F=\dd\mathcal A$. We use natural units $G=c=1$  and
\begin{equation}
     \begin{aligned}
     \dd s^2={}&-f(\dd t-\omega\dd\varphi)^2\\
     &+f^{-1}\left[e^{2k}(\dd\rho^2+\dd z^2)+\rho^2\dd\varphi^2\right],\\
     \mathcal A={}&A_t\dd t+A_\varphi\dd\varphi.
     \end{aligned}\label{eq:weyl}
\end{equation}
All functions are taken to depend on $\rho$ and $z$. We define the potentials as $\psi=2A_t$ and $\kappa=e^{-\al\phi}$, together with the magnetic and twist potentials $\chi,\epsilon$ introduced in Ref \cite{Bixano:2026ouq,Bixano:2026xum}. In particular, $C=-D\kappa/\kappa$, where $D=(\partial_\rho,\partial_z)$. Note that the sign choice for $\gamma$ in Eq. \eqref{eq:modmax} differs from that adopted in Refs \cite{BallonBordo:2020jtw,Barrientos:2024umq,Bokulic:2025usc}. Consequently, their rescaling factors translate to $e^\gamma$ in our notation, and their standard nonnegative parameter corresponds to $\gamma\leq0$ in our conventions.

In the nonfrozen sector, restricting the fields to a single real harmonic function does not necessarily fix the electromagnetic phase. A varying scalar $c(\s)$ modifies the relative electric and magnetic components of a static dyon, while a complex gravitational one-form $a(\s)$ contributes to the twist. Both effects are retained in the equations for $a(\s),b(\s),c(\s)$. In the real single-charge branch, the additional ModMax terms disappear from the reduced differential equations, although $\gamma$ remains in the normalization and the physical electromagnetic potential. For the coupled static branches, we derive an exact correspondence with Maxwell fields. These results allow us to identify the ModMax configurations associated with known Maxwell solutions without treating them as distinct geometrical families. The minimally coupled static construction of Ortaggio \cite{Ortaggio:2025sip} provides a related comparison, while the scalar-coupled correspondence is established explicitly below.

The temporal and axial LWP formulations lead to different direct electromagnetic potentials \cite{Bixano:2026duf}. We use the axial formulation to recover the Melvin--swirling solution of Barrientos et al. \cite{Barrientos:2024umq} from a logarithmic harmonic. Within the same branch, the Schwarzschild axial harmonic yields an exact black-hole exterior in the ModMax background. We first present the complete fields for an arbitrary harmonic and verify the reduced equations before considering these particular configurations. We then examine the limiting behavior of the horizon and distinguish its geometric regularity from the continuation of the constitutive fields across the null equatorial locus, since smoothness of the metric alone does not guarantee regularity of the electromagnetic sector.

Each family of solutions is presented in its own section. The opening subsection provides the integration for a generic $\s$, together with its dual and the k-quadrature $K[\s]$. The subsequent subsections select the black-hole harmonic and fix the horizon domain. They also state the ModMax identification and the Maxwell limit, including the necessary parameter and gauge choices. Alternative harmonic solutions are treated as part of the same family. The shared phase and rescaling conditions are demonstrated in Sec. \ref{sec:phase}. Previously known Maxwell fields that were already included in Ref \cite{Bixano:2026xjg} are referenced without rederiving their integrations. We keep only charged configurations that display an explicit $\gamma$ dependence, modifying the energy or the spatial harmonic alone does not constitute a distinct family.

\section{Potentials method}\label{Sec:PotentialMethod}

We adopt the three field equations and one-forms from Ref. \cite{Bixano:2026ouq}, without assuming a fixed electromagnetic phase. In what follows, products of one-forms denote contractions in the Weyl basis. On a non-null patch, the invariant relation implies
\begin{equation}
     \begin{gathered}
     e^{i\Theta}=-\frac{\bar B^{\,2}}{|B^2|},\qquad
     \cos\Theta=-\frac{\Ree(B^2)}{|B^2|},\\
     \sin\Theta=\frac{B^2-\bar B^{\,2}}{2i|B^2|},
     \end{gathered}
     \label{eq:phaseclosure}
\end{equation}
where we have used 
\begin{align}
     \mathcal{F}&=\frac{2fe^{-2k}}w
       \left(|\Imm(B)|^2-|\Ree (B)|^2\right),
     \label{eq:FBRBI}\\
     \mathcal{G}&=\frac{4fe^{-2k}}w\,\Ree (B) \cdot \Imm(B) .
     \label{eq:GBRBI}
\end{align}
The symbol $\Ree$ represents the real part, whereas $\Imm$ represents the imaginary part.
Consequently, the coefficients $w,v$ and their derivatives $\lambda=\dd\ln w,\nu=\dd v/w$ can be eliminated algebraically. Substitution in the original equations gives
\begin{equation}
     \begin{aligned}
     \frac1\rho D\!\cdot(\rho A)={}&B\bar B+A(A-\bar A),\\
     \frac1\rho D\!\cdot(\rho B)={}&-\frac12B(A-3\bar A)+C\bar B\\
     &+\frac{\tanh\gamma}{2[|B^2|-\tanh\gamma\Ree(B^2)]}\\
     &\quad\times\left\{[B^2\bar B-\Ree(B^2)B]
     D\ln\frac{B^2}{|B^2|}\right.\\
     &\hspace{16mm}\left.+C(B^2-\bar B^{\,2})(B-\bar B)\right\},\\
     \frac1\rho D\!\cdot(\rho C)={}&\frac{\al^2}{\epsilon_0}|B^2|
     \frac{\Ree(B^2)-\tanh\gamma|B^2|}
     {|B^2|-\tanh\gamma\Ree(B^2)}.
     \end{aligned}\label{eq:ABCclosed}
\end{equation}
For example, differentiating the first relation in Eq. \eqref{eq:phaseclosure} gives
\begin{equation}
     D\ln\frac{B^2}{|B^2|}
     =\frac12\left[\frac{D(B^2)}{B^2}
     -\frac{D(\bar B^{\,2})}{\bar B^{\,2}}\right].
     \label{eq:phaselog}
\end{equation}
This combines the two derivative contributions in the electromagnetic equation. The residual term, which is proportional to $C$, produces the final term inside the braces. For the scalar equation, making the same substitution similarly combines the two invariant sources into the ratio shown in Eq. \eqref{eq:ABCclosed}.

For finite real $\gamma$ and $B^2\ne0$, the denominator satisfies
\begin{equation}
     |B^2|-\tanh\gamma\Ree(B^2)
     \geq(1-|\tanh\gamma|)|B^2|>0.
     \label{eq:denominator}
\end{equation}
Therefore, this elimination does not generate any pole on the present patch. A sign flip of one invariant is permissible, the sector where both invariants vanish lies outside this splitting. The $B=0$ branch is handled separately as the scalar vacuum case. When $\gamma=0$, the additional term in the second equation drops out and the scalar source reduces to $\al^2\Ree(B^2)/\epsilon_0$, thereby reproducing the Maxwell--scalar field equations.

\subsection{Harmonic field equations}\label{sec:harmonic}

In the flat auxiliary subspace $\sigma=0$, choose a real harmonic function $\s(\rho,z)$ that is satisfying
\begin{equation}
     \frac1\rho D\!\cdot(\rho D\s)=0.
     \label{eq:harmonic}
\end{equation}
If the potentials are functions of $\s$, then the associated one-forms take the form $A=a(\s)D\s$, $B=b(\s)D\s$, and $C=c(\s)D\s$. Because $B^2=b^2|D\s|^2$, the corresponding normalized phase is $b^2/|b|^2$. As a result, the shared spatial factor drops out of Eq. \eqref{eq:ABCclosed}, yielding
\begin{equation}
     \begin{aligned}
     a'={}&|b|^2+a(a-\bar a),\\
     b'={}&-\frac12b(a-3\bar a)+c\bar b\\
     &+\frac{\tanh\gamma}{2[|b|^2-\tanh\gamma\Ree(b^2)]}\\
     &\quad\times\left\{[b^2\bar b-\Ree(b^2)b]
     \left(\ln\frac{b^2}{|b|^2}\right)'\right.\\
     &\hspace{16mm}\left.+c(b^2-\bar b^{\,2})(b-\bar b)\right\},\\
     c'={}&\frac{\al^2}{\epsilon_0}|b|^2
     \frac{\Ree(b^2)-\tanh\gamma|b|^2}
     {|b|^2-\tanh\gamma\Ree(b^2)}.
     \end{aligned}\label{eq:abc}
\end{equation}
A prime indicates $\dd/\dd\s$. The electromagnetic equation is an implicit first-order relation for $b$, since the phase derivative involves $b'$. It keeps the original coefficient instead of substituting it with a different electromagnetic quantity. On the lower branches, it is solved alongside the reconstruction equations.

The coefficients $a$ and $b$ are allowed to be complex. When the scalar is real and the harmonic is real, $c=-(\ln\kappa)'$ is also real. Any complex components of $C$ written in conjugate auxiliary coordinates must nevertheless encode this same real one-form. They therefore do not introduce a separate complex-scalar branch of Eq.\eqref{eq:action}.

The conserved geodesic energy is
\begin{equation}
     H=|a|^2-|b|^2+\frac{\epsilon_0c^2}{\al^2}
       =|a|^2-|b|^2+\epsilon_0(\phi')^2.
     \label{eq:energy}
\end{equation}
The second expression remains valid at $\al=0$, for which $c=0$, although a free scalar can still be present. To verify conservation, Eq.\eqref{eq:abc} yields $(|a|^2)'=2\Ree(a)|b|^2$ and
\begin{equation}
     (|b|^2)'=2\Ree(a)|b|^2+
     2c|b|^2\frac{\Ree(b^2)-\tanh\gamma|b|^2}
     {|b|^2-\tanh\gamma\Ree(b^2)}.
     \label{eq:bnorm}
\end{equation}
The derivative of the scalar contribution in Eq.\eqref{eq:energy} cancels the second term in Eq.\eqref{eq:bnorm}. Therefore $H'=0$. This quantity corresponds to the affine energy of the harmonic trajectory in the potentials formulation \cite{Bixano:2026duf}.

\section{Integration and identification of one-harmonic branches}\label{sec:one}

\subsection{Reconstruction, normalization, and physical charges}\label{sec:reconstruction}

We begin by expressing the reconstruction in terms of the original coefficients. Removing the constitutive coefficients via their defining relations yields (See \cite{Bixano:2026xum,Bixano:2026ouq,Bixano:2026duf})
\begin{equation}
     \begin{aligned}
     (\ln f)'={}&2\Ree a,\qquad (\ln\kappa)'=-c,\\
     A_t'={}&-e^{\gamma/2}\frac{\sqrt f}{\kappa}\Ree b
     \sqrt{\frac{|b|^2}{(\Ree b)^2+e^{2\gamma}(\Imm b)^2}},\\
     \chi'={}&2e^{3\gamma/2}\kappa\sqrt f\,\Imm b
     \sqrt{\frac{|b|^2}{(\Ree b)^2+e^{2\gamma}(\Imm b)^2}},\\
     \epsilon'={}&\psi\chi'-2f\Imm a,\qquad
     \phi'=c/\al\quad(\al\ne0).
     \end{aligned}\label{eq:reconstruct}
\end{equation}
In the opposite direction, one can reconstruct the original $b$ from these derivatives according to
{\footnotesize
\begin{equation}
     \begin{split}
     b={}&-\frac{e^{-\gamma/2}}{2\sqrt f}
     \sqrt{\frac{\kappa^2(\psi')^2+\kappa^{-2}(\chi')^2}
     {e^{2\gamma}\kappa^2(\psi')^2+\kappa^{-2}(\chi')^2}} \,(e^\gamma\kappa\psi'-i\kappa^{-1}\chi').
     \end{split}\label{eq:binverse}
\end{equation}
}
These formulas reduce to the Maxwell limit discussed in Ref. \cite{Bixano:2026xjg}. 

For a harmonic function that decays asymptotically, i.e. \(\lim_{x\rightarrow \infty} \s=0\), we may set $\s=0$ at one asymptotic boundary and adopt the normalizations $f(0)=\kappa(0)=1$ along with $A_t(0)=\chi(0)=0$. Under this convention, the normalized initial data are given by
\begin{equation}
     \begin{gathered}
     a(0)=\frac{l_1+i l_2}{L},\qquad
     c(0)=-\frac{\al d_0}{L},\\
     b(0)=\frac{e^{\gamma/2}\sqrt{q_0^2+p_0^2}}
     {L\sqrt{e^{2\gamma}q_0^2+p_0^2}}(-e^\gamma q_0+i p_0),\\
     HL^2=l_1^2+l_2^2+\epsilon_0d_0^2-e^\gamma(q_0^2+p_0^2).
     \end{gathered}\label{eq:charges}
\end{equation}
In the particular situation where $q_0=p_0=0$, one can simply set $b(0)=0$ explicitly. The constant $L>0$ fixes the overall normalization of the harmonic function. These parameters can be viewed as physical charges as long as, at the chosen end, $\s=-x^{-1}+O(x^{-2})$.

The dual harmonic and the Weyl \(k\)-quadrature are
\begin{equation}
     \begin{aligned}
     D\schdual&=-\rho(\s_{,z},-\s_{,\rho}),\\
     K_{,\rho}&=\rho(\s_{,\rho}^2-\s_{,z}^2),\qquad
     K_{,z}=2\rho\s_{,\rho}\s_{,z}.
     \end{aligned}\label{eq:dualK}
\end{equation}
The harmonic equation ensures their integrability. The remaining metric and angular-potential quadratures are
\begin{equation}
     k=HK[\s],\qquad \omega=\frac{2l_2}{L}\schdual,\qquad
     A_\varphi=-\omega A_t-\frac{p_0}{L}\schdual.
     \label{eq:quadratures}
\end{equation}
The additive integration constants are determined by choosing a symmetric local angular gauge. Taking the imaginary part of the first relation in Eq. \eqref{eq:abc} leads to $(\Imm a/f)'=0$, which fixes the twist constant $l_2/L$. In the same way, the angular electromagnetic equation sets the magnetic flux constant $p_0/L$. If one differentiates the final relation, one obtains $DA_\varphi=-\omega DA_t-(2l_2A_t+p_0)D\schdual/L$, where the twist term provides its contribution to the magnetic field. These are the original reconstruction relations, presented together with their associated conserved constants \cite{Bixano:2026ouq,Bixano:2026duf,Bixano:2026xum}.

For the normalized monopole, Eq. \eqref{eq:reconstruct} gives
\begin{equation}
     \begin{aligned}
     f&=1-\frac{2l_1}{Lx}+O(x^{-2}),&
     A_t&=-\frac{e^\gamma q_0}{Lx}+O(x^{-2}),\\
     \phi&=\frac{d_0}{Lx}+O(x^{-2}),&
     \chi&=-\frac{2e^\gamma p_0}{Lx}+O(x^{-2}).
     \end{aligned}\label{eq:asymptotics}
\end{equation}
Hence $l_1,l_2,d_0$ represent, respectively, the mass, NUT, and scalar parameters. The electric charge is defined by the flux of the constitutive displacement, rather than simply by the coefficient of the $-1/r$ term in $A_t$. Using the orientation specified in Eq. \eqref{eq:phaseclosure}, the displacement two-form is
\begin{equation}
     \kappa^2\left[
     \left(\cosh\gamma+\frac{\mathcal F\sinh\gamma}
     {\sqrt{\mathcal F^2+\mathcal G^2}}\right){}^\star F
     -\frac{\mathcal G\sinh\gamma}{\sqrt{\mathcal F^2+\mathcal G^2}}F
     \right].\label{eq:fluxform}
\end{equation}
Dividing its flux by $4\pi$ gives $q_0$, while dividing the flux of $F$ by $4\pi$ yields $p_0$. This accounts for the two rescaling factors appearing in Eq. \eqref{eq:charges} and Eq. \eqref{eq:asymptotics}.

\subsection{Prolate/oblate coordinates and the constant \texorpdfstring{$L$}{L}}\label{sec:coordinates}

In both spheroidal coordinate charts, we adopt the same positive real constant $L$. The coordinates determine the spatial harmonic function, while $H$ encodes the type-path in potential space. Both conventions appear in the metric only via the combination $k=HK[\s]$. The sign of the energy alone does not choose a coordinate chart, nor does it by itself distinguish a black hole from a wormhole.

\paragraph{The prolate coordinates are}
\begin{equation}
     \rho=L\sqrt{(x^2-1)(1-y^2)},\qquad z=Lxy,
     \label{eq:prolate}
\end{equation}
with $x>1$ and $-1<y<1$. The Schwarzwald harmonic function is
\begin{equation}
     \s=\frac12\ln\frac{x-1}{x+1},\qquad
     \schdual=Ly,\qquad
     K=\frac12\ln\frac{x^2-1}{x^2-y^2}.
     \label{eq:Schharmonic}
\end{equation}
The harmonic equation can be derived from $(x^2-1)\s_{,x}=1$. Here $\s(x)$ ranges over $-\infty<\s<0$, with $\s\to-\infty$ at the rod $x=1$, while at spatial infinity one has $\s=-x^{-1}+O(x^{-3})$. The meridional metric on the weyl subspace takes the form
\begin{equation}
     \dd\rho^2+\dd z^2=L^2(x^2-y^2)
     \left[\frac{\dd x^2}{x^2-1}+\frac{\dd y^2}{1-y^2}\right].
     \label{eq:spheroidalbase}
\end{equation}
In the spherical nonextremal cases, setting $H=1$ removes the angular prefactor in this formula. One can then use the $\s\to-\infty$ asymptotics of the integrated matter fields to decide which trajectories genuinely end on a smooth horizon.

\paragraph{The oblate coordinates and normalized monopole are}
\begin{equation}
     \begin{gathered}
     \rho=L\sqrt{(x^2+1)(1-y^2)},\qquad z=Lxy,\\
     \s=\arctan x-\frac\pi2,\qquad\schdual=Ly,\\
     K=\frac12\ln\frac{x^2+y^2}{x^2+1}.
     \end{gathered}\label{eq:oblate}
\end{equation}
Here we take $-\infty<x<\infty$ and $-\pi<\s<0$. Harmonicity follows from the identity $(x^2+1)\s_{,x}=1$. The line element of the Weyl subspace reads
\begin{equation}
     \dd\rho^2+\dd z^2=L^2(x^2+y^2)
     \left[\frac{\dd x^2}{x^2+1}+\frac{\dd y^2}{1-y^2}\right].
     \label{eq:oblatebase}
\end{equation}
When this is combined with $e^{2HK}$, one obtains a factor $(x^2+y^2)/(x^2+1)$ raised to the power $H+1$. Setting $H=-1$ cancels this contribution. Nevertheless, both endpoints of the real oblate interval must remain inside the domain of the integrated fields. In particular, a zero in an electromagnetic denominator may obstruct access to the second endpoint.

The charge relation in Eq. \eqref{eq:charges} gives, in these two spherical normalizations,
\begin{equation}
     \begin{aligned}
     H=1:\quad L^2={}&l_1^2+l_2^2+\epsilon_0d_0^2
     -e^\gamma(q_0^2+p_0^2),\\
     H=-1:\quad L^2={}&e^\gamma(q_0^2+p_0^2)
     -l_1^2-l_2^2-\epsilon_0d_0^2.
     \end{aligned}\label{eq:focalenergy}
\end{equation}
In both lines $L^2>0$. One does not make $L$ imaginary to pass from one chart to the other. Away from these normalizations, it is $HL^2$, rather than $L^2$ alone, that equals the charge combination. We retain $H$ for the conserved energy and use $h$ when writing a positive branch as $H=h^2$.

\paragraph{For a zero-energy monopole we instead use the Coulomb harmonic,}
\begin{equation}
     \begin{gathered}
     \rho=Lx\sqrt{1-y^2},\qquad z=Lxy,\qquad x>0,\\
     \s=-x^{-1},\qquad\schdual=Ly,\qquad
     K=-\frac{1-y^2}{2x^2}.
     \end{gathered}\label{eq:Coulombharmonic}
\end{equation}
Here $k=0$ since $H=0$, whereas the charge relation is merely a balancing requirement and does not determine $L$. The remaining freedom in $L$ corresponds to the harmonic normalization. This separation becomes important in the extremal limit: the focal length of the prolate may go to zero, but one must not impose $L=0$ within the prolate coordinates and then divide by it. The null branch instead requires its own integration procedure and Coulomb construction.

In all three cases, the Boyer-Linquist coordinates are written using $Lx=r-l_1$ and $y=\cos\theta$. For the prolate expressions we additionally keep
\begin{equation}
     \begin{gathered}
     Lx=r-l_1,\qquad y=\cos\theta,\\
     \Delta=(r-l_1)^2-L^2,\qquad
     \Sigma=(r-l_1)^2-L^2\cos^2\theta.
     \end{gathered}\label{eq:BL}
\end{equation}
It then follows that $\rho^2=\Delta\sin^2\theta$ and $\dd\rho^2+\dd z^2=\Sigma(\dd r^2/\Delta+\dd\theta^2)$. In contrast, the oblate radial formulas display $(r-l_1)^2+L^2$ directly. Every coordinate convention used later is to be understood as belonging to this subsection.

\section{Electromagnetic phase and the ModMax deformation}\label{sec:phase}

The absence of the electromagnetic field (\(b=0\)) and a constant electromagnetic phase (\(\mathcal{G}/\mathcal{F}\)) are different restrictions. When $b=0$, from \eqref{eq:FBRBI} we can see that $\mathcal F=0$ and ModMax contributes nothing. The vacuum and scalar-vacuum branches are those of Ref. \cite{Bixano:2026xjg}. Setting $\phi=0$ removes the scalar, whereas $\al=0$ only decouples it and still allows a free scalar. Neither operation removes ModMax from a charged configuration. We retain the charged sectors below. A nonconstant free scalar at $\al=0$ diverges on the nonextremal Schwarzschild rod, so the regular electrovacuum members have $d_0=0$.

\begin{theorem}[Constant electromagnetic phase]\label{thm:frozen}
Let $\s$ be a real harmonic function, $c(\s)\in\mathbb R$, $b(\s)\ne0$, and $\gamma\ne0$ on a connected harmonic interval. The ModMax electromagnetic phase is constant if and only if
\[
    \begin{split}
    c(b^2-\bar b^{\,2})+2e^\gamma\cosh\gamma\,(a-\bar a)\,[|b|^2-\tanh\gamma\Ree(b^2)]=0.
    \end{split}
\]
throughout that interval. In particular, real $a(\s)$ and $c(\s)=0$ freeze the phase. A nonzero real or purely imaginary $b(\s)$ requires real $a(\s)$. For real $a(\s)$, a dyon with $b^2\ne\bar b^{\,2}$ and $c\ne0$ has a varying phase.
\end{theorem}

\begin{proof}
Multiply the electromagnetic equation in Eq. \eqref{eq:abc} by $\bar b(\s)$, subtract its conjugate multiplied by $b(\s)$, and solve for the logarithmic phase derivative. This gives
\[
    \begin{split}
    \left(\ln\frac{b^2}{|b|^2}\right)'={}&
    -\frac{c(b^2-\bar b^{\,2})}{|b|^2}\\
    &-2e^\gamma\cosh\gamma\,(a-\bar a)
    \frac{|b|^2-\tanh\gamma\Ree(b^2)}{|b|^2}.
    \end{split}
\]
Since $e^{i\Theta}=-\bar b^2/|b|^2$, the electromagnetic phase is constant precisely when this derivative vanishes. Multiplication by $|b|^2>0$ proves both directions of the criterion. For real or purely imaginary $b(\s)$, its first term is zero. Moreover,
\[
    |b|^2-\tanh\gamma\Ree(b^2)
    \ge(1-|\tanh\gamma|)|b|^2>0.
\]
The second term can then vanish only if $a=\bar a$. For real $a(\s)$, the criterion instead reduces to $c(b^2-\bar b^{\,2})=0$, proving the last assertion.
\end{proof}

When $\gamma=0$, the constitutive coefficients coincide with the Maxwell constants in every phase, when $b(\s)=0$, there is no electromagnetic phase left to freeze. If $\al\ne0$, imposing $c=0$ further enforces $\Ree(b^2)=\tanh\gamma|b|^2$ via the scalar equation. The real part of the normalized phase is therefore fixed, and continuity determines its imaginary part along the connected interval. Theorem \ref{thm:frozen} then implies that $a(\s)$ must be real. With the normalization $\kappa=1$, the scalar source constraint becomes $q_0^2=p_0^2$. Hence a constant coupled scalar yields the balanced static dyon, not the general charged NUT solution. Similarly, a nonzero real or purely imaginary $b(\s)$ requires real $a(\s)$, as shown in Theorem \ref{thm:frozen}. This restriction does not exclude the charged twisting families with complex $a$ and $b$.

Substituting the phase derivative back into the electromagnetic equation makes the two ModMax terms explicit, without replacing $b$:
\[
    \begin{split}
    b'={}&-\tfrac12b(a-3\bar a)+c\bar b\\
    &+\frac{c\tanh\gamma\,b(b^2-\bar b^{\,2})^2}
    {4|b|^2[|b|^2-\tanh\gamma\Ree(b^2)]}\\
    &+\frac{e^\gamma\sinh\gamma}{2}(a-\bar a)
    \frac{b(b-\bar b)^2}{|b|^2}.
    \end{split}
\]
The first and third equations remain Eq. \eqref{eq:abc}. The family reductions below follow by imposing their stated conditions in these equations.

\begin{theorem}[Rescaling and twist]\label{thm:rescaling}
In the static harmonic sector \(a(\s)\in \mathbb{R}\), the ModMax--scalar equations for either $\epsilon_0$ and any real $\al$ have an exact Maxwell--scalar rescaling with
\begin{equation}
     q_0^{\rm M}=e^{\gamma/2}q_0,\qquad
     p_0^{\rm M}=e^{\gamma/2}p_0.
     \label{eq:staticembeddingcharges}
\end{equation}
For the same $f,\kappa,k$ and normalized additive gauges,
\[
    \psi=e^{\gamma/2}\psi_{\rm M},\qquad
    \chi=e^{\gamma/2}\chi_{\rm M},
\]
and the physical four-potentials obey
\begin{equation}
    A_t=e^{\gamma/2}A_t^{\rm M},\qquad
    A_\varphi=e^{-\gamma/2}A_\varphi^{\rm M}.
    \label{eq:staticembedding}
\end{equation}
This correspondence allows for an electromagnetic phase (\(\Theta\)) that can vary. On a charged interval $b(\s)\ne0$ with $\Imm a\ne0$ and $\gamma\ne0$, there are no nonzero real constants $C_\psi,C_\chi$ for which $\psi=C_\psi\psi_{\rm M}$ and $\chi=C_\chi\chi_{\rm M}$ map the Maxwell electromagnetic equations to ModMax while keeping the same normalized
metric, scalar field and twist.
\end{theorem}

\begin{proof}
Using Eqs. \eqref{eq:abc} and \eqref{eq:reconstruct}, the electromagnetic equation can be written as
\begin{equation}
     \begin{aligned}
     \left(\frac{\kappa\psi'}{\sqrt f}\right)'={}&
     (\Ree a+c)\frac{\kappa\psi'}{\sqrt f}
     -2e^\gamma\Imm a\frac{\chi'}{\kappa\sqrt f},\\
     \left(\frac{\chi'}{\kappa\sqrt f}\right)'={}&
     (\Ree a-c)\frac{\chi'}{\kappa\sqrt f}
     +2e^\gamma\Imm a\frac{\kappa\psi'}{\sqrt f}.
     \end{aligned}\label{eq:twistcoupling}
\end{equation}
Its Maxwell counterpart is obtained by setting $\gamma=0$.

First we take $a=\bar a$. Multiplying the first equation by $\kappa/\sqrt f$ and the second by $1/(\kappa\sqrt f)$, and using $(\ln f)'=2a$ and $(\ln\kappa)'=-c$, gives
\[
     \left(\frac{\kappa^2\psi'}f\right)'=0,\qquad
     \left(\frac{\chi'}{f\kappa^2}\right)'=0.
\]
The initial conditions in Eq. \eqref{eq:charges} therefore imply
\[
     \frac{\kappa^2\psi'}f=\frac{2e^\gamma q_0}{L},
     \qquad
     \frac{\chi'}{f\kappa^2}=\frac{2e^\gamma p_0}{L}.
\]
These are the static first integrals in the potential formulation of Refs. \cite{Bixano:2026ouq,Bixano:2026duf}, expressed in the normalization used here.

To compare the gravitational and scalar equations, substitution of Eq. \eqref{eq:binverse} gives
\[
     \begin{aligned}
     |b|^2&=\frac{e^{-\gamma}}{4f}
     \left[\kappa^2(\psi')^2+\kappa^{-2}(\chi')^2\right],\\
     c'&=\frac{\al^2e^{-\gamma}}{4\epsilon_0f}
     \left[\kappa^2(\psi')^2-\kappa^{-2}(\chi')^2\right].
     \end{aligned}
\]
Since $a'=|b|^2$ in the static sector and $\psi=2A_t$, the first integrals yield
\begin{equation}
     \begin{aligned}
     A_t'&=\frac{e^\gamma q_0f}{L\kappa^2},\qquad
     \chi'=\frac{2e^\gamma p_0f\kappa^2}{L},\\
     a'&=\frac{e^\gamma f}{L^2}
           (q_0^2\kappa^{-2}+p_0^2\kappa^2),\\
     c'&=\frac{\al^2e^\gamma f}{\epsilon_0L^2}
           (q_0^2\kappa^{-2}-p_0^2\kappa^2).
     \end{aligned}\label{eq:staticproof}
\end{equation}

Define $q_0^{\rm M}=e^{\gamma/2}q_0$ and $p_0^{\rm M}=e^{\gamma/2}p_0$. With this choice, the final two relations coincide precisely with the Maxwell--scalar equations carrying charges $q_0^{\rm M}$ and $p_0^{\rm M}$. Along with $f'=2af$ and $\kappa'=-c\kappa$, they yield the same initial-value problem for $f,a,\kappa,$ and $c$. Hence, identical initial conditions imply identical solutions on their shared interval of regularity. When $\al\neq 0$, the scalar field is then also the same via $\kappa=e^{-\al\phi}$. For $\al=0$, the free-scalar equation remains the same, and one separately keeps the same initial data for the scalar.

For these common profiles, the electromagnetic first integrals give
\[
     \psi'=e^{\gamma/2}\psi_{\rm M}',\qquad
     \chi'=e^{\gamma/2}\chi_{\rm M}'.
\]
The normalized additive gauges therefore lead to $\psi=e^{\gamma/2}\psi_{\rm M}$ and $\chi=e^{\gamma/2}\chi_{\rm M}$. It follows that $A_t=e^{\gamma/2}A_t^{\rm M}$. From \eqref{eq:charges}, one obtains the same $H$ after replacing the charges, and thus the harmonic reconstruction yields the identical $k=HK$ \cite{Bixano:2026duf}. Lastly, taking $\omega=0$,
\[
     A_\varphi=-\frac{p_0}{L}\schdual
     =e^{-\gamma/2}
     \left(-\frac{p_0^{\rm M}}{L}\schdual\right)
     =e^{-\gamma/2}A_\varphi^{\rm M}.
\]
This proves the static correspondence without imposing a constant electromagnetic phase.

Now take a charged interval with $\Imm a\ne0$ and assume
\[
    \psi=C_\psi\psi_{\rm M},\qquad \chi=C_\chi\chi_{\rm M},
\]
with nonzero real $C_\psi,C_\chi$, and the same $f,\kappa,a,c$ in both theories. Substituting into Eq. \eqref{eq:twistcoupling} and subtracting the Maxwell equations scaled by $C_\psi$ and $C_\chi$ yields, since $f,\kappa,\Imm a\neq0$,
\[
    (C_\psi-e^\gamma C_\chi)\chi_{\rm M}'=0,\qquad
    (C_\chi-e^\gamma C_\psi)\psi_{\rm M}'=0.
\]
Neither $\psi_{\rm M}'$ nor $\chi_{\rm M}'$ can vanish identically: if one did, the Maxwell equations (with $\Imm a\ne0$) force the other to vanish, making the rescaled electromagnetic field zero, contradicting that the interval is charged. Hence
\[
    C_\psi=e^\gamma C_\chi,\qquad C_\chi=e^\gamma C_\psi,
\]
so $(1-e^{2\gamma})C_\psi=0$. With $C_\psi\ne0$ and real $\gamma$, this implies $\gamma=0$, contradicting a nonzero ModMax deformation. Therefore no diagonal constant rescaling exists at fixed metric, scalar, and twist on a charged interval.
\end{proof}

The second theorem rules out diagonal constant rescalings for the chosen metric and scalar, it should not be read as a full classification of every solution-generating transformation. Here we establish its static component while keeping the scalar source in place, and Ortaggio’s minimally coupled extension criterion \cite{Ortaggio:2025sip} offers a pertinent point of comparison. Consequently, a nonconstant phase by itself does not amount to a genuinely new geometry. Below, we separate the explicit static embeddings from the nonfrozen twisting solutions.

\paragraph{The common horizon test.}
The spatial normalization follows Ref. \cite{Bixano:2026xjg}. First impose finite matter fields as $\s\to-\infty$. If this gives $f\propto e^{2h\s}$ and $H=h^2$, then
\begin{equation}
    \begin{split}
    \sqrt{g_{yy}g_{\varphi\varphi}}
    &\sim\frac{L^2}{f}(x^2-1)^{(H+1)/2}
    (x^2-y^2)^{(1-H)/2}\\
    &\propto(x-1)^{(h-1)^2/2}.
    \end{split}\label{eq:rodareadensity}
\end{equation}
For the area to be finite and nonzero, we must have $h=1$, and therefore $H=1$. While this requirement is necessary, each family must additionally yield a finite scalar and maintain positive denominators for all $r\ge r_+=l_1+L$. The degenerate Coulomb limits, however, call for a separate check of the area. NUT entries correspond to local Killing horizons, with their standard, separately defined axis identification. We use $\dd\Omega_2^2=\dd\theta^2+\sin^2\theta\dd\varphi^2$.

\section{Static electrovacuum and the RN family}\label{sec:RN}

\subsection{Integration for an arbitrary harmonic}\label{sec:generalRN}

Let $\al=0$ and $\phi(\s)=0$, so $c(\s)=0$. Imposing $a=\bar a$ while keeping $b(\s)$ complex. The field equations \eqref{eq:abc} reduces to
\begin{equation}
    a'=|b|^2,\qquad b'=ab,\qquad c=0,
    \qquad H=a^2-|b|^2.
    \label{eq:RNsystem}
\end{equation}
Both extra terms in $b'$ vanish: the scalar coefficient is zero and the twist coefficient $a-\bar a$ is zero. Hence $\gamma$ drops out of the reduced differential equation, but remains in the initial charge normalization in Eq. \eqref{eq:charges} and in the reconstructed electric potential.

Eliminating $|b|^2$ gives $a'=a^2-H$. With $a=-\mathcal D'/\mathcal D$, this becomes $\mathcal D''=H\mathcal D$, while $b'=ab$ gives $b=b(0)/\mathcal D$. The initial values $\mathcal D(0)=1$, $\mathcal D'(0)=-l_1/L$ therefore gives
\begin{equation}
     \begin{aligned}
     \mathcal D&=\cosh(h\s)-\frac{l_1}{hL}\sinh(h\s),\\
     a&=-\frac{\mathcal D'}{\mathcal D},\qquad
     b=\frac{b(0)}{\mathcal D},\qquad c=0,\\
     h^2L^2&=l_1^2-e^\gamma(q_0^2+p_0^2).
     \end{aligned}\label{eq:RNabc}
\end{equation}

For \(h>0\), we choose the positive branch of the denominator interval that includes \(\s=0\). A first integral is given by \(\mathcal D'^2-h^2\mathcal D^2=\frac{e^\gamma(q_0^2+p_0^2)}{L^2}\).
Consequently, \(a'=|b|^2\) remains valid using the original complex \(b\) and with electromagnetic charge taken to be independent. Using the identity \([\sinh(h\s)/(h\mathcal D)]'=\mathcal D^{-2}\), the electric field can be integrated. The six physical functions are
\begin{equation}
    \begin{aligned}
    f&=\mathcal D^{-2},\qquad \omega=0,\qquad k=HK,\\
    A_t&=\frac{e^\gamma q_0\sinh(h\s)}{hL\mathcal D},\qquad
    A_\varphi=-\frac{p_0}{L}\schdual,\qquad\phi=0.
    \end{aligned}\label{eq:RNharmonic}
\end{equation}
This is the integration for an arbitrary harmonic function. At $h=0$, its analytic limit is $\mathcal D=1-l_1\s/L$ and $A_t=e^\gamma q_0\s/(L\mathcal D)$.

\subsection{Spacetime solutions and identification}

\subsubsection{The Schwarzschild harmonic and its horizons}

For the Schwarzschild harmonic function \eqref{eq:Schharmonic}, using the BL coordinates \eqref{eq:BL}, and $\mathcal D\propto e^{-h\s}$ near the candidate horizon. Equation \eqref{eq:rodareadensity} requires $h=1$. Then $\mathcal D=r/\sqrt\Delta$, and the solution takes the form
\begin{equation}
     \begin{aligned}
     f&=1-\frac{2l_1}{r}+\frac{e^\gamma(q_0^2+p_0^2)}{r^2},\\
     \dd s^2&=-f\dd t^2+f^{-1}\dd r^2+r^2\dd\Omega_2^2,\\
     \mathcal A&=-\frac{e^\gamma q_0}{r}\dd t-p_0\cos\theta\dd\varphi,
     \qquad\phi=0,\\
     L^2&=l_1^2-e^\gamma(q_0^2+p_0^2)>0.
     \end{aligned}\label{eq:RN}
\end{equation}

The charged, nonextremal regime is defined by $l_1>0$ and $0<e^\gamma(q_0^2+p_0^2)<l_1^2$. The denominator $r$ stays positive for all $r\ge r_+=l_1+L$. The horizon area and the surface gravity are given by $4\pi r_+^2$ and $L/r_+^2$, respectively. In the linear-limit case, the Coulomb harmonic yields $f=(1-l_1/r)^2$ with $l_1^2=e^\gamma(q_0^2+p_0^2)$, implying that the degenerate horizon has area $4\pi l_1^2$.

\paragraph{ModMax identification.}
\textbf{Known ModMax solution.} Equation \eqref{eq:RN} corresponds to the dyonic Reissner--Nordstr\"om--ModMax black hole discussed by Flores-Alfonso et al. \cite{Flores-Alfonso:2020euz} and by Amirabi--Mazharimousavi \cite{Amirabi:2020mzv}. In their conventions, the parameter with the standard sign is $-\gamma$, the electric parameter is the displacement charge $q_0$, and the magnetic parameter is $p_0$. With these identifications, one reproduces both the squared charge appearing in the metric and the four-potential, modulo an additive electric gauge term.

\paragraph{Maxwell limit.}
Setting $\gamma=0$ yields the RN branch, its degenerate Coulomb case, and the charged Weyl solutions of Ref. \cite{Bixano:2026xjg}. The multi-centre limit reduces to Maxwell Majumdar--Papapetrou \cite{Majumdar:1947eu,Papaetrou:1947ib}, and the zero-intercept limit to Bertotti--Robinson.

\subsubsection{Other harmonics: multiple horizons and the product limit}

For ordered rod endpoints $z_A$, put $R_A=\sqrt{\rho^2+(z-z_A)^2}$ and take $\sigma_A=-1,+1$ at the lower and upper endpoint of each rod. Then
\[
     \begin{aligned}
     \s&=\frac12\sum_A\sigma_A\ln[R_A+z-z_A],\\
     \schdual&=-\frac12\sum_A\sigma_A R_A,\\
     K&=-\frac14\sum_{A,B}\sigma_A\sigma_B
     \ln\frac{R_AR_B+(z-z_A)(z-z_B)+\rho^2}
     {2R_AR_B}.
     \end{aligned}
\]

Taking derivatives of the first line yields $\s_{,\rho}=-\sum_A\sigma_A(z-z_A)/(2\rho R_A)$ and $\s_{,z}=\sum_A\sigma_A/(2R_A)$. Inserting these into Eq. \eqref{eq:dualK} produces the quoted expression for $K$, with the cross terms arising from distinct rods included. It is precisely these mixed terms that fix the conical forces, keeping only the single-rod pieces would not suffice.

\textbf{Known ModMax construction.} Substitute these functions into Eq. \eqref{eq:RNharmonic}. In this way one recovers the multiple-black-hole fields with a common charge ratio found by Bokuli\'c--Herdeiro in \cite{Bokulic:2025usc}. To match their conventions, set $V=h\s$, $e^{2U}=f$, $\gamma_{\rm ref}=-\gamma$, take the electric potential as $\chi_{\rm ref}=-A_t$, and identify the charging ratio as $q_{\rm ref}=q_0/l_1$. Using Eq. \eqref{eq:RNabc}, their square-root factor translates to $hL/l_1$. In general, the nonextremal portions of the axis include struts.

The balanced limit uses $H=0$ and
\[
     \mathcal D=1+\sum_i\frac{l_{1i}}
     {\sqrt{\rho^2+(z-z_i)^2}},\qquad l_{1i}>0.
\]

With $l_1=\sum_i l_{1i}$ and $\s=L(1-\mathcal D)/l_1$, the complete fields are
\[
     \begin{aligned}
     \dd s^2={}&-\mathcal D^{-2}\dd t^2+
     \mathcal D^2(\dd\rho^2+\dd z^2+\rho^2\dd\varphi^2),\\
     A_t={}&\frac{e^\gamma q_0}{l_1}(\mathcal D^{-1}-1),\qquad
     A_\varphi=-\frac{p_0}{L}\schdual,\qquad\phi=0,\\
     l_1^2={}&e^\gamma(q_0^2+p_0^2).
     \end{aligned}
\]

Positive residues give the known ModMax Majumdar--Papapetrou horizons \cite{Bokulic:2025usc,Majumdar:1947eu,Papaetrou:1947ib}. Their individual charges are $q_{0i}=q_0l_{1i}/l_1$ and $p_{0i}=p_0l_{1i}/l_1$. For this harmonic the dual is explicitly
\[
    \schdual=\frac{L}{l_1}\sum_i
    \frac{l_{1i}(z-z_i)}{\sqrt{\rho^2+(z-z_i)^2}},
\]
up to an angular gauge parameter. If the centres are collinear, the axisymmetric ansatz is maintained. Choosing multiple centres alters how the horizon is realized, but leaves the reduced system \eqref{eq:RNsystem} unchanged.

\paragraph{The near-horizon product.}
The zero-intercept version of the same null denominator gives the ModMax Bertotti--Robinson product,
\[
     \begin{aligned}
     \dd s^2={}&-\frac{r^2}{e^\gamma(q_0^2+p_0^2)}\dd t^2
     +e^\gamma(q_0^2+p_0^2)\frac{\dd r^2}{r^2}\\
     &+e^\gamma(q_0^2+p_0^2)\dd\Omega_2^2,\\
     \mathcal A={}&\frac{q_0r}{q_0^2+p_0^2}\dd t
     -p_0\cos\theta\dd\varphi,\qquad\phi=0.
     \end{aligned}
\]
The squares of its two radii are $e^\gamma(q_0^2+p_0^2)$. Equivalently, one can obtain this from the near-horizon limit of the extremal fields in Eq. \eqref{eq:RN}, once the constant electric gauge potential is removed. This product is already known \cite{Kubiznak:2026arb}, it should be viewed as a limit within the RN family, not as a separate charged black-hole solution.

\section{The complex electrovacuum branch}\label{sec:NUT}

\subsection{Integration for an arbitrary harmonic}\label{sec:generalNUT}

We keep $\al=0$ and $\phi(\s)=0$, but now permit both $a(\s)$ and $b(\s)$ to be complex, while setting $c(\s)=0$. The resulting reduced equations are
\[
    \begin{aligned}
    a'&=|b|^2+a(a-\bar a),\qquad c=0,\\
    b'&=-\tfrac12b(a-3\bar a)
    +\frac{e^\gamma\sinh\gamma}{2}(a-\bar a)
    \frac{b(b-\bar b)^2}{|b|^2},\\
    H&=|a|^2-|b|^2.
    \end{aligned}
\]
The last term in $b'$ represents the explicit ModMax twist contribution. It does not appear in Eq. \eqref{eq:RNsystem} but remains present here when $\gamma\ne0$ and the charged twist is nonvanishing. Eliminating $|b|^2$ once more yields $a'=a^2-H$. The initial derivative of the denominator is now complex, therefore
\begin{equation}
     \begin{aligned}
     \mathcal D&=\cosh(h\s)-\frac{l_1+i l_2}{hL}\sinh(h\s),\\
     a&=-\frac{\mathcal D'}{\mathcal D},\qquad f=|\mathcal D|^{-2},\\
     |b|^2&=\frac{e^\gamma(q_0^2+p_0^2)}{L^2|\mathcal D|^2},\qquad c=0,\\
     h^2L^2&=l_1^2+l_2^2-e^\gamma(q_0^2+p_0^2).
     \end{aligned}\label{eq:NUTabc}
\end{equation}

The electromagnetic equation fixes the original coefficient's phase through
\[
     \frac{\Ree b+i e^\gamma\Imm b}
     {\sqrt{(\Ree b)^2+e^{2\gamma}(\Imm b)^2}}
     =\frac{-q_0+i p_0}{\sqrt{q_0^2+p_0^2}}
     \left(\frac{\mathcal D}{\bar{\mathcal D}}\right)^{e^\gamma}.
\]
Together with its modulus in Eq. \eqref{eq:NUTabc}, this completely determines $b$. In terms of the electromagnetic potentials, the same equation becomes as the following.
\[
    \psi'+i\chi'=
    \frac{2e^\gamma(q_0+i p_0)}{L|\mathcal D|^2}
    \left(\frac{\bar{\mathcal D}}{\mathcal D}\right)^{e^\gamma}.
\]

The derivative of $(\bar{\mathcal D}/\mathcal D)^{e^\gamma}$ is $2ie^\gamma l_2(\bar{\mathcal D}/\mathcal D)^{e^\gamma}/(L|\mathcal D|^2)$. Hence the integral, normalized at zero, is
\begin{equation}
    \psi+i\chi=\frac{q_0+i p_0}{i l_2}
    \left[\left(\frac{\bar{\mathcal D}}{\mathcal D}\right)^{e^\gamma}-1\right].
    \label{eq:NUTpotentials}
\end{equation}

Along with 
\[
\begin{aligned}
    f=|\mathcal D|^{-2}, \qquad \omega=2l_2\schdual/L, \qquad k=HK, \\
    A_t=\psi/2, \qquad A_\varphi=-\omega A_t-p_0\schdual/L , \qquad \phi=0,
\end{aligned}
\]
one obtains the complete set of six physical functions. The complex parameter $b$ is then determined from Eq. \eqref{eq:binverse} using these explicit potentials. The power is analytically continued starting from $\mathcal D(0)=1$, rather than being selected independently at each point.

For $H=0$ the same arbitrary-harmonic integration has
\[
    \mathcal D=1-\frac{l_1+il_2}{L}\s,\qquad
    e^\gamma(q_0^2+p_0^2)=l_1^2+l_2^2.
\]
The primitive in Eq. \eqref{eq:NUTpotentials} and the angular fields remain the same. Therefore, the null limit is applied to the integrated family prior to selecting any spatial configuration.

\subsubsection{Magnetic LWP representation of the same branch}\label{sec:axialgeneral}

For $\al=0$ and $\phi=0$, the complex electrovacuum integration likewise admits a magnetic LWP representation. Its denominator satisfies the same linear equation as Eq. \eqref{eq:NUTabc}, however, switching the Killing reduction flips the sign of the electromagnetic norm and modifies the constitutive phase factor. Adopting the axial convention of Ref. \cite{Bixano:2026duf}, we write
\begin{equation}
    \begin{split}
    \dd s^2={}&f_{\rm m}(\dd\varphi-\omega_{\rm m}\dd t)^2\\
    &+f_{\rm m}^{-1}\left[e^{2k_{\rm m}}(\dd\rho^2+\dd z^2)-\rho^2\dd t^2\right].
    \end{split}\label{eq:axialmetric}
\end{equation}
Here $\psi_m=2A_\varphi$. With background amplitudes $E_{\rm B},B_{\rm B}$ and swirl $j$, the denominator is
\[
    \mathcal D_{\rm m}=L^{-1}e^{-\s}
    +L\left[\frac{e^\gamma(E_{\rm B}^2+B_{\rm B}^2)}4+ij\right]e^\s.
\]
It satisfies $\mathcal D_{\rm m}''=\mathcal D_{\rm m}$ and $|\mathcal D_{\rm m}'|^2-|\mathcal D_{\rm m}|^2=-e^\gamma(E_{\rm B}^2+B_{\rm B}^2)$. Thus the same denominator integration gives
\begin{equation}
    \begin{aligned}
    V&=1+\frac{e^\gamma(E_{\rm B}^2+B_{\rm B}^2)}4L^2e^{2\s},\\
    F_{\rm B}&=V^2+j^2L^4e^{4\s},\qquad
    f_{\rm m}=\frac{L^2e^{2\s}}{F_{\rm B}}.
    \end{aligned}\label{eq:axialfactors}
\end{equation}
The angular field is the corresponding phase primitive, with an axial factor $e^{-\gamma}$:
\begin{equation}
    \begin{split}
    A_\varphi={}&\frac{e^\gamma B_{\rm B}}{4j}
    \sin\left[2e^{-\gamma}\arctan\frac{jL^2e^{2\s}}{V}\right]\\
    &+\frac{e^\gamma E_{\rm B}}{4j}
    \left\{\cos\left[2e^{-\gamma}\arctan\frac{jL^2e^{2\s}}{V}\right]-1\right\}.
    \end{split}\label{eq:axialAphi}
\end{equation}
The other fields follow from the same dual and $K$ reconstructions,
\[
    \begin{aligned}
    \omega_{\rm m}&=-4j\schdual,\qquad e^{2k_{\rm m}}=L^2e^{2K},\\
    A_t&=(e^\gamma E_{\rm B}+4jA_\varphi)\schdual,\qquad\phi=0.
    \end{aligned}
\]

The factor $V$ satisfies $V'=2(V-1)$. Consequently,
\[
     \frac{\dd}{\dd\s}\arctan\frac{jL^2e^{2\s}}{V}
     =\frac{2jL^2e^{2\s}}{V^2+j^2L^4e^{4\s}}
     =2jf_{\rm m}.
\]
The complete sine/cosine phase in Eq. \eqref{eq:axialAphi} thus has derivative $4je^{-\gamma}f_{\rm m}$. Taking two derivatives of $A_\varphi$ yields
\[
     \left(\frac{A_\varphi'}{f_{\rm m}}\right)'
     =-4je^{-2\gamma}f_{\rm m}
           (e^\gamma E_{\rm B}+4jA_\varphi).
\]
Multiplying through by $e^\gamma$ yields the first identity stated below. To obtain the second identity, note that the formulas for $A_\varphi'/f_{\rm m}$ and $E_{\rm B}+4je^{-\gamma}A_\varphi$ represent a sine/cosine rotation of the pair $(B_{\rm B},\,E_{\rm B})$. When these rotated expressions are squared and summed, the cross terms cancel, leaving $E_{\rm B}^2+B_{\rm B}^2$. This confirms the electromagnetic equation as well as the quadratic source combination employed in the reconstruction.

\[
    \begin{aligned}
    &\left(\frac{e^\gamma A_\varphi'}{f_{\rm m}}\right)'
    +4je^{-\gamma}f_{\rm m}(e^\gamma E_{\rm B}+4jA_\varphi)=0,\\
    &e^\gamma(A_\varphi')^2+
    e^{-\gamma}f_{\rm m}^2(e^\gamma E_{\rm B}+4jA_\varphi)^2\\
    &\hspace{15mm}=e^\gamma(E_{\rm B}^2+B_{\rm B}^2)f_{\rm m}^2.
    \end{aligned}
\]
The $j=0$ limit is $A_\varphi=B_{\rm B}L^2e^{2\s}/(2V)$ and $A_t=e^\gamma E_{\rm B}\schdual$. The temporal metric functions, when that chart has $f>0$, are
\begin{equation}
    \begin{aligned}
    f&=\frac{\rho^2}{f_{\rm m}}-f_{\rm m}\omega_{\rm m}^2,\qquad
    \omega=-\frac{f_{\rm m}\omega_{\rm m}}f,\\
    e^{2k}&=\frac{f}{f_{\rm m}}e^{2k_{\rm m}}.
    \end{aligned}\label{eq:axialtotemporal}
\end{equation}
These relationships arise immediately from the Killing and meridional blocks. The expressions provide an alternative harmonic and LWP description of this electrovacuum branch, not a separate integration over matter.

\subsection{Spacetime representations and identification}

\subsubsection{Schwarzschild harmonic in the electric LWP frame}

The same finite-area reasoning implies $h=1$. Inserting the Schwarzschild harmonic gives $\mathcal D=(r+il_2)/\sqrt\Delta$ and
\[
     \begin{aligned}
     f={}&\frac{\Delta}{r^2+l_2^2},\\
     \dd s^2={}&-f(\dd t-2l_2\cos\theta\dd\varphi)^2
     +f^{-1}\dd r^2+(r^2+l_2^2)\dd\Omega_2^2,\\
     A_t={}&\frac1{2l_2}\Imm\left\{(q_0+i p_0)
     \left[\left(\frac{r-i l_2}{r+i l_2}\right)^{e^\gamma}-1\right]\right\},\\
     \mathcal A={}&A_t(\dd t-2l_2\cos\theta\dd\varphi)
     -p_0\cos\theta\dd\varphi,\qquad\phi=0.
     \end{aligned}
\]

In this notation $L^2=l_1^2+l_2^2-e^\gamma(q_0^2+p_0^2)>0$. When $l_1\ge0$, the denominator stays positive for $r>r_+=l_1+L$. The (local) horizon has area $4\pi(r_+^2+l_2^2)$ and surface gravity $L/(r_+^2+l_2^2)$. The standard NUT-axis identification is still a separate global consideration.

In the integrated $H=0$ member, the Coulomb harmonic function substitution gives
\[
     \mathcal D=\frac{r+i l_2}{r-l_1},\qquad
     f=\frac{(r-l_1)^2}{r^2+l_2^2}.
\]

For $l_1>0$ its area is $4\pi(l_1^2+l_2^2)$. This is the known degenerate NUT limit. 

\paragraph{ModMax identification.}
\textbf{Known ModMax solution.} These coincide with the fields presented by Ballon Bordo, Kubiz\v{n}\'ak, and Perche in \emph{Taub--NUT solutions in conformal electrodynamics} \cite{BallonBordo:2020jtw}. In that paper they take $\Lambda=0$, with the identifications $m=l_1$, $n=l_2$, $e=q_0$, $g=-p_0/(2l_2)$, and $\gamma_{\rm ref}=-\gamma$. One must also flip the angular coordinate, $\varphi_{\rm ref}=-\varphi$, and choose $a_{\rm ref}=A_t+p_0/(2l_2)$. With these conventions, the four-potentials differ only by the term $p_0\dd t/(2l_2)$. Moreover, their phase $2e^{-\gamma_{\rm ref}}\arctan(n/r)$ matches precisely the argument of our complex power.

\paragraph{Maxwell limit.}
For $\gamma=0$, the exponent in Eq. \eqref{eq:NUTpotentials} equals unity. The resulting configurations coincide with the Maxwell-charged NUT branch reported in Ref. \cite{Bixano:2026xjg}. Its degenerate charged member, with $H=0$ and $q_0^2+p_0^2=l_1^2+l_2^2$, belongs to the one-real-harmonic IWP sector \cite{Israel:1972vx,Perjes:1971gv}. The full IWP family permits independent real and imaginary harmonic data and extends beyond the present ansatz.

Accordingly, these established Maxwell fields do not require an additional integration in our treatment.

\subsubsection{Logarithmic harmonic: the Barrientos background}\label{sec:logrepresentation}

Taking
\[
    \s=\ln(\rho/L),\qquad \schdual=z,\qquad K=\ln(\rho/L).
\]
The relations in Eq. \eqref{eq:dualK} are a direct consequence of $\rho\s_{,\rho}=1$. It follows that $L^2e^{2\s}=\rho^2$ and $e^{2k_{\rm m}}=\rho^2$, so Eq. \eqref{eq:axialmetric} can be rewritten as
\[
    \dd s^2=\frac{\rho^2}{F_{\rm B}}(\dd\varphi+4jz\dd t)^2
    +F_{\rm B}(-\dd t^2+\dd\rho^2+\dd z^2).
\]

\paragraph{ModMax/Maxwell identification.}
\textbf{Known ModMax solution.} This corresponds to the Melvin--Bonnor--swirling background discussed by Barrientos et al. in Sec. V of Ref. \cite{Barrientos:2024umq}, taking $\gamma_{\rm ref}=-\gamma$, $E_{\rm ref}=E_{\rm B}$, $B_{\rm ref}=B_{\rm B}$, and $j_{\rm ref}=j$. On the continuous axial branch, their angle $\pi-2\arctan[V/(j\rho^2)]$ is equivalent to $2\arctan(j\rho^2/V)$, so the two potential components match, including the term $4jzA_\varphi$. For $\gamma=0$ one recovers the familiar Maxwell electromagnetic swirling universe \cite{Barrientos:2024pkt}. Thus, the harmonic and LWP formulation provides the correct identification, rather than introducing an additional branch.

\subsubsection{Axial Schwarzschild harmonic and its exact exterior}\label{sec:swirlBH}

For $l_1=L>0$, taking the Schwarzschild harmonic function in the axial LWP frame:
\begin{equation}
    \begin{aligned}
    \s_{\rm m}&=\ln(\rho/L)-\s_{\rm Sch}
    =\ln\frac{r\sin\theta}{L},\\
    \widetilde{\s}_{\rm m}&=z-L\cos\theta
    =(r-2l_1)\cos\theta,\\
    K[\s_{\rm m}]&=\ln(\rho/L)+K[\s_{\rm Sch}]-2\s_{\rm Sch}.
    \end{aligned}\label{eq:swirlBHharmonic}
\end{equation}
The harmonic and its dual follow by linearity from Eq. \eqref{eq:Schharmonic}. The function $K$ is quadratic, so its mixed term must also be included. Substitution in Eq. \eqref{eq:dualK} gives
\[
    \begin{aligned}
    K[\s_{\rm m}]_{,\rho}
    &=\rho^{-1}+K[\s_{\rm Sch}]_{,\rho}-2\s_{{\rm Sch},\rho},\\
    K[\s_{\rm m}]_{,z}
    &=K[\s_{\rm Sch}]_{,z}-2\s_{{\rm Sch},z}.
    \end{aligned}
\]
Integrating these relations, with the same additive normalization, gives the last line of Eq. \eqref{eq:swirlBHharmonic}.

We now substitute $L^2e^{2\s_{\rm m}}=r^2\sin^2\theta$ in Eqs. \eqref{eq:axialfactors}--\eqref{eq:axialAphi}.
The metric and the remaining fields become
\[
    \begin{aligned}
    \dd s^2={}&F_{\rm B}\left[
    -\left(1-\frac{2l_1}{r}\right)\dd t^2
    +\frac{\dd r^2}{1-2l_1/r}+r^2\dd\theta^2\right]\\
    &+\frac{r^2\sin^2\theta}{F_{\rm B}}
    [\dd\varphi+4j(r-2l_1)\cos\theta\,\dd t]^2,\\
    A_t={}&(e^\gamma E_{\rm B}+4jA_\varphi)(r-2l_1)\cos\theta,
    \qquad\phi=0.
    \end{aligned}
\]
Here $V,F_{\rm B}$ and $A_\varphi$ are the same functions already integrated above, evaluated at this harmonic. The electric LWP functions follow from Eq. \eqref{eq:axialtotemporal}.

For finite real parameters, $V\ge1$ and $F_{\rm B}\ge1$. Thus there is no denominator zero for $r\ge2l_1$.
At $r_+=2l_1$, the mixed angular coefficient vanishes, and the horizon generator is $\partial_t$. Moreover,
\[
    \begin{aligned}
    \sqrt{g_{\theta\theta}g_{\varphi\varphi}}\big|_{r_+}
    &=r_+^2\sin\theta,\\
    \operatorname{Area}_+&=16\pi l_1^2,
    \qquad\varkappa_+=\frac1{4l_1}.
    \end{aligned}
\]
The cancellation of $F_{\rm B}$ in the areas element explains why the background does not change the horizon area. The surface gravity follows from one half of the radial derivative of $1-2l_1/r$ at $r_+$.

To check the fields at the horizon, we set $\dd v=\dd t+\dd r/(1-2l_1/r)$. The angular one-form becomes
\[
    \dd\varphi+4j(r-2l_1)\cos\theta\,\dd v
    -4jr\cos\theta\,\dd r.
\]
The radial coefficient of the four-potential is then $-(e^\gamma E_{\rm B}+4jA_\varphi)r\cos\theta$, which is finite.
Hence the metric, the potential and $F=\dd\mathcal A$ have smooth limit from the exterior. On the axes, $F_{\rm B}=1$ and
$A_\varphi=O(\sin^2\theta)$, so an angular period $2\pi$ gives elementary flatness. Also, $g^{tt}=-[F_{\rm B}(1-2l_1/r)]^{-1}<0$ in the exterior. The spacetime is not asymptotically flat, so $l_1$ is used as the seed horizon scale rather than being assigned an ADM mass.

Electromagnetic regularity requires a further check. Direct contraction of $F$ gives
\[
    \begin{split}
    \sqrt{\mathcal F^2+\mathcal G^2}
    ={}&\frac{2|1-(2l_1/r)\sin^2\theta|}{F_{\rm B}^2}\\
    &\times\left[
    \left(\frac{A_\varphi'}{f_{\rm m}}\right)^2
    +(e^\gamma E_{\rm B}+4jA_\varphi)^2\right].
    \end{split}
\]
The prime here is with respect to $\s_{\rm m}$. Both invariants therefore vanish at $r=2l_1$, $\theta=\pi/2$. This does not mean that the electromagnetic tensor vanishes. In the ingoing coordinates, its value on that circle is
\[
    F=\dd r\wedge\left[
    \frac{A_\varphi'}{2l_1}\dd\varphi
    -2l_1(e^\gamma E_{\rm B}+4jA_\varphi)\dd\theta\right].
\]
The quadratic identity obtainde above reads
\[
    \left(\frac{A_\varphi'}{f_{\rm m}}\right)^2
    +e^{-2\gamma}(e^\gamma E_{\rm B}+4jA_\varphi)^2
    =E_{\rm B}^2+B_{\rm B}^2.
\]
For a nonzero electromagnetic background, the two coefficients cannot both vanish. The horizon circle therefore carries a finite, nonzero null field.

The ratios in Eq. \eqref{eq:fluxform} have finite exterior limits: the common spatial factor cancels, and the remaining functions depend smoothly on $\s_{\rm m}$. For $\gamma\ne0$, the invariant constitutive relation becomes $0/0$ at the circle. Thus ModMax is not differentiable there in terms of the two invariants, though null fields can be handled in the Hamiltonian formulation \cite{Bandos:2020jsw}. The factor $1-(2l_1/r)\sin^2\theta$ changes sign across $r=2l_1\sin^2\theta$, so the normalized invariants must be recomputed using the positive square root on the other side. Hence smooth $g$ and $F$ do not by themselves define a constitutive continuation across the horizon. We claim only an exact exterior solution with finite horizon limits, any two-sided constitutive extension must be checked separately. For $\gamma=0$, the invariant ratios drop out of the field equations.

\paragraph{ModMax identification.}
\textbf{Known ModMax solution for $j=0$.}
Taking $m=l_1$, $E=E_{\rm B}$, $B=B_{\rm B}$ and $\gamma_{\rm ref}=-\gamma$ gives the Schwarzschild--Melvin--Bonnor fields
of Sec. IV of Ref. \cite{Barrientos:2024umq}:
\[
    A_t=e^\gamma E_{\rm B}(r-2l_1)\cos\theta,
    \qquad
    A_\varphi=\frac{B_{\rm B}r^2\sin^2\theta}{2V}.
\]
Taking the zero-mass limit of the reconstructed fields gives the known background of Sec. \ref{sec:logrepresentation}. This limit is taken after reconstruction, since the harmonic normalization above assumes $L=l_1>0$.

\textbf{New ModMax family,  .}
For $j\gamma\ne0$ and $E_{\rm B}^2+B_{\rm B}^2>0$, the construction gives the Schwarzschild--Melvin--swirling exterior left open in Ref. \cite{Barrientos:2024umq}. It is a spacetime representation of the same axial harmonic branch. Its ModMax dependence includes the factor $e^{-\gamma}$ inside the electromagnetic phase in Eq. \eqref{eq:axialAphi}, not only the amplitude $e^\gamma(E_{\rm B}^2+B_{\rm B}^2)$ in $V$. At fixed metric, constant amplitude rescalings cannot remove this change of phase along the harmonic trajectory. The horizon statement is subject to the constitutive qualification just established.

\paragraph{Maxwell limit and recent comparisons.}
At $\gamma=0$, the full fields agree with Sec. 3.4 of Ref. \cite{Barrientos:2024pkt}, using $M=l_1$, $x=\cos\theta$, $j_{\rm ref}=j$, $E_{\rm ref}=-E_{\rm B}$, $B_{\rm ref}=-B_{\rm B}$ and $|X|^2=(E_{\rm B}^2+B_{\rm B}^2)/4$.

For $E_{\rm B}=0$, they also follow from Eqs. (3.1)--(3.8) of Di Pinto, Klemm and Vigan\`o \cite{DiPinto:2025yaa} by taking
\[
    a=Q=H=0,\qquad M=l_1,\qquad
    B=B_{\rm B},\qquad\jmath=j.
\]
Their $H$ denotes magnetic charge, not our geodesic energy. Their functions then satisfy $\Sigma=r^4$, $F=r^2F_{\rm B}$, $A_0=0$ and
\[
    \frac{\Omega}{\Sigma}=-4j(r-2l_1)\cos\theta,\qquad
    \frac{A_3}{F}=\frac{B_{\rm B}r^2\sin^2\theta\,V}{2F_{\rm B}}.
\]
Thus both the metric and the four-potential coincide.

Astorino \cite{Astorino:2026kuv} places the Maxwell background in a larger family with an additional curling parameter $\aleph$. In our zero-mass, purely magnetic Maxwell limit, his Eq. (2.15) agrees with the metric for $b=B_{\rm B}$, $\jmath=-j$ and $\Delta_\phi=1$. The potential agrees with the Harrison representative in his footnote 8, his Eq. (2.16) is related to that representative by electromagnetic duality, not by a gauge transformation. The curling parameter is absent from our fields. In particular, his Schwarzschild solution in Sec. 3.1 has curling rather than the independent swirling parameter used here, so it is not the same general black-hole configuration.

\section{A single charge and a coupled scalar}\label{sec:GM}

\subsection{Integration for an arbitrary harmonic}\label{sec:generalGM}

We choose $\al=1,\sqrt3$. The opposite sign can be recovered by applying $\al\to-\al$, $\phi\to-\phi$, and $d_0\to-d_0$, while keeping $\kappa$ unchanged. The special phantom case $\epsilon_0=-1$, $\al=1$ (Super String Low Energies (SS--LE) theory) is included in the same family, because the general expression for the denominator contains a division by $1+\epsilon_0\al^2$.

In the electric sector, set $a=\bar a$ and $b=\bar b$, while keeping $c$ real and dynamical. Because $a-\bar a=b-\bar b=b^2-\bar b^{\,2}=0$, the complete system simplifies precisely to
\[
     a'=b^2,\qquad b'=(a+c)b,\qquad
     c'=\frac{\al^2}{\epsilon_0}b^2.
\]

The additional contributions to $b'$ drop out. ModMax still appears in $b(0)=-e^{\gamma/2}q_0/L$ and in the relation converting $b$ into the physical electric field.

Subtracting $\epsilon_0\al^2a'$ from $c'$ yields the constant $c-\epsilon_0\al^2a=-\al(d_0+\epsilon_0\al l_1)/L$. If we instead add the equations, we obtain $(a+c)'=(1+\epsilon_0\al^2)b^2$. Combined with $b'=(a+c)b$, it follows that $(a+c)^2-(1+\epsilon_0\al^2)b^2=\lambda^2$ remains constant. Therefore $a+c=-\mathcal D_e'/\mathcal D_e$ again satisfies a linear denominator equation. The initial conditions determine
\[
     \begin{aligned}
     \lambda^2&=\frac{(l_1-\al d_0)^2
     -(1+\epsilon_0\al^2)e^\gamma q_0^2}{L^2},\\
     \mathcal D_e&=\cosh(\lambda\s)
     -\frac{l_1-\al d_0}{\lambda L}\sinh(\lambda\s),\\
     a&=\frac{-\mathcal D_e'/\mathcal D_e+
     \al(d_0+\epsilon_0\al l_1)/L}{1+\epsilon_0\al^2},\\
     b&=-\frac{e^{\gamma/2}q_0}{L\mathcal D_e},\qquad
     c=\epsilon_0\al^2a-\frac{\al(d_0+\epsilon_0\al l_1)}{L}.
     \end{aligned}
\]

Integrating $(\ln f)'=2a$ and $\phi'=c/\al$ then gives
\[
     \begin{aligned}
     f&=\exp\!\left[\frac{2\al(d_0+\epsilon_0\al l_1)\s}
     {(1+\epsilon_0\al^2)L}\right]
     \mathcal D_e^{-2/(1+\epsilon_0\al^2)},\\
     \phi&=-\frac{\epsilon_0\al}{1+\epsilon_0\al^2}\ln\mathcal D_e
     -\frac{d_0+\epsilon_0\al l_1}{(1+\epsilon_0\al^2)L}\s,\\
     A_t&=\frac{e^\gamma q_0\sinh(\lambda\s)}
     {\lambda L\mathcal D_e},\\
     A_\varphi&=0,\qquad\omega=0,\qquad k=HK,\\
     H&=\frac{l_1^2+\epsilon_0d_0^2-e^\gamma q_0^2}{L^2}\\
     &=\frac{\lambda^2+
     \epsilon_0(d_0+\epsilon_0\al l_1)^2/L^2}
     {1+\epsilon_0\al^2}.
     \end{aligned}
\]

The electric integral is obtained from $f/\kappa^2=\mathcal D_e^{-2}$ together with the same derivative identity as in the RN case. In the magnetic sector, $b$ is instead purely imaginary, and the same relations simplify to
\[
    a'=|b|^2,\qquad b'=(a-c)b,\qquad
    c'=-\frac{\al^2}{\epsilon_0}|b|^2.
\]
The full field configuration can be read off from the electric formulas by making the replacements $q_0\to p_0$, $d_0\to-d_0$, and $\phi\to-\phi$, which leaves the resulting $f$ and $k$ unaffected. For the gauge potential, taking $A_t=0$ and $A_\varphi=-p_0\schdual/L$ in place of the electric potential. With these substitutions, Eq. \eqref{eq:binverse} yields $b(0)=ie^{\gamma/2}p_0/L$. In either sector, the single-charge phase remains fixed.

The standard zero-energy limit follows from the same harmonic forms, prior to choosing a monopole:
\[
     \begin{gathered}
     \mathcal D_e=1-\frac{(1+\al^2)l_1}{L}\s,\qquad
     f=\mathcal D_e^{-2/(1+\al^2)},\\
     \phi=-\frac{\al}{1+\al^2}\ln\mathcal D_e,
     \qquad A_t=\frac{e^\gamma q_0\s}{L\mathcal D_e}.
     \end{gathered}
\]
In this case one has $d_0=-\al l_1$ and $e^\gamma q_0^2=(1+\al^2)l_1^2$. Applying the same reconstruction procedure yields $\omega=k=A_\varphi=0$.

\subsubsection{The exceptional SS--LE phantom case}

Fixing $\epsilon_0=-1$, $\al=1$, the reduced equations are
\[
     a'=b^2,\qquad b'=(a+c)b,\qquad c'=-b^2.
\]

Now $a+c=\lambda=(l_1-d_0)/L$ is constant. First integrate $b'=\lambda b$, then $a'=b^2$ and $c=\lambda-a$. Integrating $f'/f=2a$ and $\phi'=c$ gives
\[
     \begin{aligned}
     b={}&-\frac{e^{\gamma/2}q_0}{L}e^{\lambda\s},\\
     a={}&\frac{l_1}{L}+\frac{e^\gamma q_0^2}{2\lambda L^2}
     (e^{2\lambda\s}-1),\qquad c=\lambda-a,\\
     f={}&\exp\left[\frac{2l_1}{L}\s+
     \frac{e^\gamma q_0^2}{2\lambda^2L^2}
     (e^{2\lambda\s}-1-2\lambda\s)\right],\\
     \phi={}&\lambda\s-\frac12\ln f,\qquad
     A_t=\frac{e^\gamma q_0}{2\lambda L}(e^{2\lambda\s}-1),\\
     \omega={}&0,\qquad A_\varphi=0,\qquad k=HK,\\
     H={}&\frac{l_1^2-d_0^2-e^\gamma q_0^2}{L^2}.
     \end{aligned}
\]

\subsection{The Schwarzschild harmonic and the black-hole domains}

In the charged, nonextremal case, one finds $\mathcal D_e\propto e^{-\lambda\s}$ at the prospective horizon. The term multiplying $\s$ in the scalar vanishes exactly if $d_0+\epsilon_0\al l_1=\epsilon_0\al\lambda L$. Under this requirement, the horizon limits become $c\to0$, $a\to\lambda$, and $H=\lambda^2$. The area criterion then determines
\begin{equation}
     \begin{gathered}
     \lambda=1,\qquad
     d_0=\epsilon_0\al(L-l_1),\\
     L^2=l_1^2+\epsilon_0d_0^2-e^\gamma q_0^2>0.
     \end{gathered}
     \label{eq:GMregularity}
\end{equation}

Only after imposing these conditions do we insert the Schwarzschild harmonic function \eqref{eq:Schharmonic}. The solution become
\begin{equation}
     \begin{aligned}
     f={}&\frac{r-l_1-L}{r-\al d_0}
     \left(\frac{r-l_1+L}{r-\al d_0}\right)^{
     (1-\epsilon_0\al^2)/(1+\epsilon_0\al^2)},\\
     \dd s^2={}&-f\dd t^2+f^{-1}\dd r^2+
     \frac{\Delta}{f}\dd\Omega_2^2,\\
     A_t={}&-\frac{e^\gamma q_0}{r-\al d_0},\qquad A_\varphi=0,\\
     \phi={}&\frac{\epsilon_0\al}{1+\epsilon_0\al^2}
     \ln\frac{r-l_1+L}{r-\al d_0}.
     \end{aligned}\label{eq:GMradial}
\end{equation}

An extra condition in the exterior region is that $r-\al d_0>0$ for every $r\ge r_+=l_1+L$. Because this term grows as $r$ increases, requiring it to be positive on the horizon rules out any scalar pole outside the horizon. A straightforward computation of $4\pi\Delta/f$ and of $f'(r_+)/2$ yields
\[
     \begin{aligned}
     \operatorname{Area}_+={}&4\pi(l_1+L-\al d_0)^{2/(1+\epsilon_0\al^2)}\\
     &\quad\times(2L)^{2\epsilon_0\al^2/(1+\epsilon_0\al^2)},\\
     \varkappa_+={}&\frac1{2(l_1+L-\al d_0)}\\
     &\quad\times\left(\frac{2L}{l_1+L-\al d_0}\right)^{
     (1-\epsilon_0\al^2)/(1+\epsilon_0\al^2)}.
     \end{aligned}
\]

For standard zero-energy configurations, the Coulomb centre has vanishing area and the scalar diverges when $\al\ne0$. Hence, while the integrated limit is well-defined, it does not correspond to a smooth black-hole endpoint with finite area.

For $\epsilon_0=-1$, $\al^2=3$, the same horizon calculation instead gives
\begin{equation}
 \begin{gathered}
 \frac{l_1}{2}<L<l_1,\qquad
 d_0=\sqrt3(l_1-L),\\
 e^\gamma q_0^2=2(l_1-L)(2L-l_1).
 \end{gathered}
 \label{eq:GMphantomBH}
\end{equation}

Over the interval shown, the charge product remains positive. It reaches its peak value $e^\gamma q_0^2=l_1^2/4$ at $L=3l_1/4$. The lower boundary is not included since $r-\al d_0$ goes to zero at the horizon. The two allowed values of $L$ to the left of this maximum correspond to distinct scalar charges within the same class. They represent two phantom horizon regions associated with the same static charge replacement. Their identification is presented below.

\subsubsection{The exceptional SS--LE phantom horizon}\label{sec:phantomunit}

To keep the scalar finite and nonextremal at the horizon, the linear term in its asymptotic expansion must disappear. This leads to $l_1/L-e^\gamma q_0^2/(2\lambda L^2)=\lambda$. The energy is then $\lambda^2$, and the area constraint therefore yields
\[
     \begin{gathered}
     \lambda=1,\qquad d_0=l_1-L,\\
     e^\gamma q_0^2=2L(l_1-L),\qquad 0<L<l_1.
     \end{gathered}
\]

Substituting the harmonic yields
\begin{equation}
     \begin{aligned}
     f&=\frac{r-l_1-L}{r-l_1+L}
     \exp\left[-\frac{2d_0}{r-l_1+L}\right],\\
     A_t&=-\frac{e^\gamma q_0}{r-l_1+L},\qquad A_\varphi=0,\\
     \phi&=\frac{d_0}{r-l_1+L},\\
     \dd s^2&=-f\dd t^2+f^{-1}\dd r^2+
     \frac{\Delta}{f}\dd\Omega_2^2.
     \end{aligned}\label{eq:phantomunitBH}
\end{equation}

The positive denominator is at least $2L$ throughout the exterior. The area and surface gravity are
\[
 \operatorname{Area}_+=16\pi L^2e^{d_0/L},
 \qquad \varkappa_+=\frac{e^{-d_0/L}}{4L}.
\]

The charge bound is $e^\gamma q_0^2\le l_1^2/2$, attained at $L=l_1/2$. The charged $\lambda=0$ integration has $H=-e^\gamma q_0^2/L^2$ and does not meet these horizon conditions

\subsubsection{ModMax identification}

\textbf{Known ModMax single-charge construction.} When $\epsilon_0=+1$, choosing the range $l_1>L>0$ yields the Gibbons--Maeda black hole, including the unit-coupling GHS limit. The corresponding Maxwell dictionary is
\begin{equation}
     \begin{gathered}
     r_{\rm GM}=r-\al d_0,\quad
     r_{\pm,\rm GM}=l_1\pm L-\al d_0,\\
     Q_{\rm GM}=e^{\gamma/2}q_0,\qquad
     r_{+,\rm GM}r_{-,\rm GM}=(1+\al^2)e^\gamma q_0^2.
     \end{gathered}\label{eq:GMdictionary}
\end{equation}

This yields $f=(1-r_{+,\rm GM}/r_{\rm GM})(1-r_{-,\rm GM}/r_{\rm GM})^{(1-\al^2)/(1+\al^2)}$, using the usual spherical radius. These solutions correspond to the Gibbons--Maeda \cite{Gibbons:1987ps} and Garfinkle--Horowitz--Strominger \cite{Garfinkle:1990qj} black holes. The physical electric potential is obtained by multiplying their Maxwell potential by $e^{\gamma/2}$. The ModMax single-charge construction is described by the sectorial maps in Ref. \cite{Bixano:2026duf}, and the associated extremal fields are written out explicitly in \cite{Bokulic:2025ucy}. 

For $\epsilon_0=-1$ and $\al^2=3$, the Maxwell profile belongs to the phantom EMD class found by Cl\'ement--Fabris--Rodrigues \cite{Clement:2009ai}. In their conventions the charge is $e^{\gamma/2}q_0$, and to match our normalization their electric potential must be rescaled by a factor of $e^{\gamma/2}$.

\textbf{New explicit ModMax solution.} The Maxwell black-hole profile has already been obtained in Cl\'ement--Fabris--Rodrigues, \emph{Phantom black holes in Einstein--Maxwell--dilaton theory} \cite{Clement:2009ai}. Starting from their Eqs. (3.26)--(3.28), identify the usual radial coordinate as $r-l_1+L$, the horizon radius as $2L$, and the exponential parameter as $d_0/L$. One then recovers Eq. \eqref{eq:phantomunitBH} at $\gamma=0$. For $\gamma\neq 0$, applying the substitutions $q_0^{\rm M}=e^{\gamma/2}q_0$ and $A_t=e^{\gamma/2}A_t^{\rm M}$ yields the complete ModMax solution. This specific ModMax branch is a new extension.

\subsubsection{Maxwell limit}

Choosing $\gamma=0$ yields the Gibbons--Maeda branch and the magnetic GHS branch analyzed in Ref. \cite{Bixano:2026xjg}, with the radial correspondence specified in Eq. \eqref{eq:GMdictionary}. The phantom regimes, including the special unit case, match the findings of Ref. \cite{Clement:2009ai}. This atypical solution does not fall within the standard-scalar classification. The novel aspect here is providing its explicit ModMax extension.

\section{Static dyons at SS-LE theory}\label{sec:string}

\subsection{Integration for an arbitrary harmonic function}\label{sec:generalstring}

For static dyons, we set $a=\bar a$, taking $c\in\mathbb R$, and assume both flux electromagnetic charges are nonzero. Plugging in the phase derivative from Theorem \ref{thm:frozen} yields
\begin{equation}
    \begin{aligned}
    a'={}&|b|^2,\\
    b'={}&ab+c\bar b+
    \frac{c\tanh\gamma\,b(b^2-\bar b^{\,2})^2}
    {4|b|^2[|b|^2-\tanh\gamma\Ree(b^2)]},\\
    c'={}&\frac{\al^2}{\epsilon_0}|b|^2
    \frac{\Ree(b^2)-\tanh\gamma|b|^2}
    {|b|^2-\tanh\gamma\Ree(b^2)}.
    \end{aligned}\label{eq:staticgamma}
\end{equation}
Substituting $\psi'=2e^\gamma q_0f/(L\kappa^2)$ and $\chi'=2e^\gamma p_0f\kappa^2/L$ from Eq. \eqref{eq:staticproof} into Eq. \eqref{eq:binverse} yields the original complex coefficient
\begin{equation}
     \begin{split}
     b={}&\frac{e^{\gamma/2}\sqrt f}{L}
     \sqrt{\frac{q_0^2\kappa^{-2}+p_0^2\kappa^2}
     {e^{2\gamma}q_0^2\kappa^{-2}+p_0^2\kappa^2}}\,(-e^\gamma q_0\kappa^{-1}+ip_0\kappa).
     \end{split}\label{eq:staticb}
\end{equation}
Taking the modulus squared then gives
\[
 |b|^2=\frac{e^\gamma f}{L^2} (q_0^2\kappa^{-2}+p_0^2\kappa^2)=a'.
\]
The phase obeys
\[
     \begin{split}
     \left(\ln\frac{b^2}{|b|^2}\right)'
     &=\frac{4ie^\gamma q_0p_0c}
     {e^{2\gamma}q_0^2\kappa^{-2}+p_0^2\kappa^2}\\
     &=-\frac{c(b^2-\bar b^{\,2})}{|b|^2},
     \end{split}
\]
which matches exactly the requirement in Theorem \ref{thm:frozen} for real $a$. Hence the phase is non-constant whenever $q_0p_0c\ne0$, even though the metric and scalar continue to satisfy the static correspondence stated in Theorem \ref{thm:rescaling}.

Choosing $\epsilon_0=+1$ and setting $\al=1$ in Eq. \eqref{eq:staticgamma}, the electromagnetic equation is unaffected. If we take the sum and the difference of the gravitational and scalar equations in Eq. \eqref{eq:staticproof}, we obtain
\[
    \begin{aligned}
    (a+c)'&=\frac{2e^\gamma q_0^2}{L^2}\frac f{\kappa^2},\\
    (a-c)'&=\frac{2e^\gamma p_0^2}{L^2}f\kappa^2,\\
    (\ln(\sqrt f/\kappa))'&=a+c,\qquad
    (\ln(\sqrt f\kappa))'=a-c.
    \end{aligned}
\]
In each case, the corresponding pair reduces to the same Riccati-type integration as in Eq. \eqref{eq:RNsystem}. Consequently, the two normalized denominators are
\begin{equation}
     \begin{aligned}
     h_e^2&=\frac{(l_1-d_0)^2-2e^\gamma q_0^2}{L^2},\\
     h_m^2&=\frac{(l_1+d_0)^2-2e^\gamma p_0^2}{L^2},\\
     \mathcal D_e&=\cosh(h_e\s)-\frac{l_1-d_0}{h_eL}\sinh(h_e\s),\\
     \mathcal D_m&=\cosh(h_m\s)-\frac{l_1+d_0}{h_mL}\sinh(h_m\s).
     \end{aligned}\label{eq:stringdenominators}
\end{equation}

We use $a=-\tfrac12(\ln\mathcal D_e\mathcal D_m)'$, $c=-\tfrac12[\ln(\mathcal D_e/\mathcal D_m)]'$, and $H=(h_e^2+h_m^2)/2$, while Eq. \eqref{eq:staticb} determines $b$. Upon integrating the corresponding logarithmic derivatives together with the electric first integral, one obtains
\[
     \begin{aligned}
     f&=(\mathcal D_e\mathcal D_m)^{-1},\qquad
     \kappa=\sqrt{\mathcal D_e/\mathcal D_m},\\
     \dd s^2&=-f\dd t^2+f^{-1}
     \left[e^{2HK}(\dd\rho^2+\dd z^2)+\rho^2\dd\varphi^2\right],\\
     \mathcal A&=\frac{e^\gamma q_0\sinh(h_e\s)}{h_eL\mathcal D_e}\dd t
     -\frac{p_0}{L}\schdual\dd\varphi,\\
     \phi&=\tfrac12\ln(\mathcal D_m/\mathcal D_e).
     \end{aligned}
\]

To derive the zero-energy case, we take the limit $\sinh(h\s)/h \to \s$ in both the denominator and the antiderivative. Taking the same limit simultaneously for the same arbitrary-harmonic solution yields
\[
     \begin{aligned}
     l_1&=\frac{e^{\gamma/2}}{\sqrt2}(|q_0|+|p_0|),\\
     d_0&=\frac{e^{\gamma/2}}{\sqrt2}(|p_0|-|q_0|),\\
     \mathcal D_e&=1-\frac{l_1-d_0}{L}\s,\qquad
     \mathcal D_m=1-\frac{l_1+d_0}{L}\s.
     \end{aligned}
\]

\subsection{The Schwarzschild harmonic and the black-hole domain}

We now select the Schwarzwald harmonic function \eqref{eq:Schharmonic}. Near its prospective horizon, the scalar scales behaves as $(h_e-h_m)\s/2$. Requiring it to remain finite forces $h_e=h_m=h$. The energy therefore becomes $h^2$, and imposing the finite-area condition sets $h=1$. Taken together, these two constraints imply
\begin{equation}
     2l_1d_0=e^\gamma(p_0^2-q_0^2),\qquad
     L^2=l_1^2+d_0^2-e^\gamma(q_0^2+p_0^2)>0.
     \label{eq:stringcharges}
\end{equation}

Reconstruction now yields
\begin{equation}
     \begin{aligned}
     f&=\frac{\Delta}{r^2-d_0^2},\\
     \dd s^2&=-f\dd t^2+f^{-1}\dd r^2+(r^2-d_0^2)\dd\Omega_2^2,\\
     \mathcal A&=-\frac{e^\gamma q_0}{r-d_0}\dd t
     -p_0\cos\theta\dd\varphi,\qquad
     \phi=\tfrac12\ln\frac{r+d_0}{r-d_0}.
     \end{aligned}\label{eq:stringdyon}
\end{equation}

The full exterior region is defined by $r\ge r_+=l_1+L>|d_0|$. If we solve the first charge constraint for $d_0$ and the constant $L^2$, we obtain
\[
     \begin{split}
     L^2={}&\frac{[2l_1^2-e^\gamma(|q_0|+|p_0|)^2]}{4l_1^2}\,[2l_1^2-e^\gamma(|q_0|-|p_0|)^2].
     \end{split}
\]

For $l_1>0$, the regular charged branch is
\[
     l_1>\frac{e^{\gamma/2}}{\sqrt2}(|q_0|+|p_0|).
\]

For two nonzero charges, the other positive \(L^2\) range is \(0<l_1<\frac{e^{\gamma/2}}{\sqrt2}\big||q_0|-|p_0|\big|\), which exists only if \(|q_0|\ne|p_0|\). The scalar-charge relation implies \(|d_0|>l_1\). Also, \(L^2-(|d_0|-l_1)^2=-2e^\gamma\min(q_0^2,p_0^2)<0\), so with \(L>0\) and \(|d_0|-l_1>0\) we get \(r_+=l_1+L<|d_0|\). Thus the surface \(r=|d_0|\) lies outside the candidate horizon. Indeed,
\[
\phi_{,r}=-\frac{d_0}{r^2-d_0^2},\qquad (\nabla\phi)^2=\frac{\Delta d_0^2}{(r^2-d_0^2)^3},
\]
and \(\Delta(|d_0|)=2e^\gamma\min(q_0^2,p_0^2)>0\), so the invariant diverges at \(r=|d_0|\). Hence this range yields a naked singularity. In the regular range \(r_+>|d_0|\),
\[
\operatorname{Area}_+=4\pi(r_+^2-d_0^2),\qquad \varkappa_+=\frac12 f_{,r}(r_+)=\frac{L}{r_+^2-d_0^2}.
\]

For the simultaneous linear limit described above, we take the Coulomb harmonic and the same radial fields \eqref{eq:stringdyon}, setting $\Delta=(r-l_1)^2$. When both charges are nonzero, one obtains
\[
     \operatorname{Area}_H=8\pi e^\gamma|q_0p_0|,\qquad
     \phi_H=\frac12\ln\left|\frac{p_0}{q_0}\right|.
\]

This result is obtained by computing $r^2-d_0^2$ and the logarithmic term at $r=l_1$. If either charge is removed, the area vanishes and the scalar diverges, so the regular degenerate horizon is no longer maintained.

\paragraph{ModMax identification}
\textbf{New explicit ModMax extension.}
The Maxwell seed is the unit-coupling EMD dyon of Ref. \cite{Bixano:2026xjg}. Theorem \ref{thm:rescaling}, with $q_0^{\rm M}=e^{\gamma/2}q_0$ and $p_0^{\rm M}=e^{\gamma/2}p_0$, gives Eq. \eqref{eq:stringdyon} and its scalar-charge relations.
The metric and scalar retaining the Maxwell profiles, while
\[
    A_t=e^{\gamma/2}A_t^{\rm M},\qquad
    A_\varphi=e^{-\gamma/2}A_\varphi^{\rm M}.
\]
For $q_0p_0d_0\ne0$, both parts of $b$ and $c$ are nonzero. Since $a$ is real, Theorem \ref{thm:frozen} then implies a varying electromagnetic phase. Thus this is an explicit two-charge, nonfrozen ModMax extension of a known Maxwell geometry. It is not a new integration of the static gravitational equations. The one-charge solutions of Ref. \cite{Bokulic:2025ucy} and the balanced RN limit $q_0^2=p_0^2$, $d_0=0$, are already known and are excluded from this novelty claim.

\paragraph{Maxwell limit}
Choosing $\gamma=0$ reproduces the static unit-coupling dyon and the associated scalar-charge constraint reported in Ref. \cite{Bixano:2026xjg}. The correspondence between charges and potentials was already provided in Eqs. \eqref{eq:staticembeddingcharges}--\eqref{eq:staticembedding}. Moreover, its Maxwell metric and the finite-area endpoint carrying two charges are established results, and do not represent new findings arising from the present deformation.

\section{Static dyons at Kaluza--Klein theory}\label{sec:KK}

\subsection{Integration for an arbitrary harmonic function}\label{sec:generalKK}

Taking $\epsilon_0=+1$ and $\al^2=3$ within the same reduced system \eqref{eq:staticgamma}. In particular, one has
\[
    a'=|b|^2,\qquad
    c'=3|b|^2\frac{\Ree(b^2)-\tanh\gamma\,|b|^2}
    {|b|^2-\tanh\gamma\,\Ree(b^2)},
\]
whereas $b'$ is determined by the $\gamma$-dependent electromagnetic equation appearing in Eq. \eqref{eq:staticgamma}. Moreover, because $[\ln(f\kappa^{-2})]'=2(a+c)$ and $[\ln(f\kappa^2)]'=2(a-c)$, it follows from Eq. \eqref{eq:staticproof} that
\begin{equation}
     \begin{aligned}
     (\ln(f\kappa^{-2}))''&=\frac{8e^\gamma q_0^2}{L^2}f\kappa^{-2}
     -\frac{4e^\gamma p_0^2}{L^2}f\kappa^2,\\
     (\ln(f\kappa^2))''&=-\frac{4e^\gamma q_0^2}{L^2}f\kappa^{-2}
     +\frac{8e^\gamma p_0^2}{L^2}f\kappa^2.
     \end{aligned}\label{eq:KKtoda}
\end{equation}

We solve the polynomial black-hole geodesic for this Toda system. The scalar charge is bounded from above as described below. Consider two quadratic factors in $e^{2h\s}$ whose linear terms are $-2\delta_e$ and $-2\delta_m$. Enforcing consistency with the bilinear equations sets their shared quadratic coefficient to $\delta_e\delta_m$:
\[
     \begin{aligned}
     \mathcal P_e&=1-2\delta_e e^{2h\s}+\delta_e\delta_m e^{4h\s},\\
     \mathcal P_m&=1-2\delta_m e^{2h\s}+\delta_e\delta_m e^{4h\s}.
     \end{aligned}
\]

For instance, differentiating directly yields \( (\ln\mathcal P_e)''=-8h^2\delta_e e^{2h\s} \frac{\mathcal P_m}{\mathcal P_e^2} \), together with the corresponding electric--magnetic relation. These expressions satisfy Eq. \eqref{eq:KKtoda} with
\[
     \begin{aligned}
     f&=e^{2h\s}\sqrt{\frac{\mathcal P_e(0)\mathcal P_m(0)}
     {\mathcal P_e\mathcal P_m}},\\
     \kappa&=\left[\frac{\mathcal P_e\mathcal P_m(0)}
     {\mathcal P_m\mathcal P_e(0)}\right]^{3/4},\\
     a&=h-\tfrac14(\ln\mathcal P_e\mathcal P_m)',\qquad
     c=-\tfrac34\left(\ln\frac{\mathcal P_e}{\mathcal P_m}\right)'.
     \end{aligned}
\]

Matching the coefficients fixes the electromagnetic charges:
\begin{equation}
     \frac{e^\gamma q_0^2}{L^2}=
     \frac{2h^2\delta_e\mathcal P_m(0)}{\mathcal P_e(0)^2},\qquad
     \frac{e^\gamma p_0^2}{L^2}=
     \frac{2h^2\delta_m\mathcal P_e(0)}{\mathcal P_m(0)^2}.
     \label{eq:KKemcharges}
\end{equation}

The normalized derivatives at zero then determine the mass and scalar charge,
\[
     \begin{aligned}
     \frac{l_1}{L}={}&h+
     \frac{h\delta_e(1-\delta_m)}{\mathcal P_e(0)}+
     \frac{h\delta_m(1-\delta_e)}{\mathcal P_m(0)},\\
     \frac{d_0}{L}={}&-
     \frac{\sqrt3h\delta_e(1-\delta_m)}{\mathcal P_e(0)}+
     \frac{\sqrt3h\delta_m(1-\delta_e)}{\mathcal P_m(0)}.
     \end{aligned}
\]

, and eliminating the polynomial parameters gives the following.
\begin{equation}
     \begin{split}
     2d_0(d_0^2-3l_1^2)={}&3e^\gamma q_0^2(d_0+\sqrt3l_1)\\
     &+3e^\gamma p_0^2(d_0-\sqrt3l_1),
     \end{split}\label{eq:KKconstraint}
\end{equation}

The energy takes the form $H=h^2=[l_1^2+d_0^2-e^\gamma(q_0^2+p_0^2)]/L^2$. Hence, the cubic serves as a constraint along this geodesic rather than representing independent scalar hair. The electric first integral is straightforward, since the bilinear identity also determines the derivative of $(1-\delta_e e^{2h\s})/\mathcal P_e$. Its normalized primitive is
\[
     A_t=\frac{hL}{q_0}\left[
     \frac{1-\delta_e e^{2h\s}}{\mathcal P_e}
     -\frac{1-\delta_e}{\mathcal P_e(0)}\right],\qquad q_0\ne0.
\]

Taking a derivative and applying Eq. \eqref{eq:KKemcharges} yields precisely $A_t'=e^\gamma q_0f/(L\kappa^2)$. The other physical functions are $\omega=0$, $k=h^2K$, $A_\varphi=-p_0\schdual/L$, and
\[
    \phi=-\frac{\sqrt3}{4}\ln
    \frac{\mathcal P_e\mathcal P_m(0)}{\mathcal P_m\mathcal P_e(0)}.
\]

Setting $H=0$ and solving the resulting cubic yields
\[
     \begin{aligned}
     l_1={}&\frac{e^{\gamma/2}}2
     (|q_0|^{2/3}+|p_0|^{2/3})^{3/2},\\
     d_0={}&\frac{\sqrt3e^{\gamma/2}}2
     (|p_0|^{2/3}-|q_0|^{2/3})
     (|q_0|^{2/3}+|p_0|^{2/3})^{1/2}.
     \end{aligned}
\]

The two finite normalized factors are
\begin{equation}
    \begin{aligned}
    \mathcal D_e={}&1-\frac{2(l_1-d_0/\sqrt3)}L\s
    +\frac{2e^\gamma|q_0|^{4/3}|p_0|^{2/3}}{L^2}\s^2,\\
    \mathcal D_m={}&1-\frac{2(l_1+d_0/\sqrt3)}L\s
    +\frac{2e^\gamma|p_0|^{4/3}|q_0|^{2/3}}{L^2}\s^2.
    \end{aligned}\label{eq:KKextremefactors}
\end{equation}
Here $f=(\mathcal D_e\mathcal D_m)^{-1/2}$, $\kappa=(\mathcal D_e/\mathcal D_m)^{3/4}$, $\phi=-\sqrt3\ln(\mathcal D_e/\mathcal D_m)/4$, $\omega=k=0$, $A_\varphi=-p_0\schdual/L$, and
\begin{equation}
    A_t=\frac{e^\gamma q_0\s}{L\mathcal D_e}
    \left[1-\frac{l_1+d_0/\sqrt3}{L}\s\right].
    \label{eq:KKextremeAt}
\end{equation}

\subsection{The Schwarzschild harmonic and the black-hole domain}

We now take the Schwarzschild harmonic function, as $\s\to-\infty$, $\mathcal P_e$ and $\mathcal P_m$ tend to one, so the scalar is finite and $f\propto e^{2h\s}$. Hence, Eq. \eqref{eq:rodareadensity} gives $h=1$. Substitute $e^{2\s}=(r-l_1-L)/(r-l_1+L)$ into all functions. The resulting radial metric reads
\[
     \begin{aligned}
     \dd s^2&=-f\dd t^2+f^{-1}\dd r^2+\frac{\Delta}{f}\dd\Omega_2^2,\\
     \mathcal A&=A_t\dd t-p_0\cos\theta\dd\varphi,
     \end{aligned}
\]

The replacement is written out directly: $\sinh(2\s)=-2L(r-l_1)/\Delta$ and $\cosh(2\s)-1=2L^2/\Delta$. Consequently, reduce to the radial polynomials 
\begin{equation}
    \begin{aligned}
    \Delta\frac{e^{-2\s}\mathcal P_e}{\mathcal P_e(0)}
    ={}&r^2-\frac{2d_0r}{\sqrt3}-\frac{d_0^2}{3}\\
    &-\frac{2l_1d_0}{\sqrt3}+e^\gamma(p_0^2-q_0^2),\\
    \Delta\frac{e^{-2\s}\mathcal P_m}{\mathcal P_m(0)}
    ={}&r^2+\frac{2d_0r}{\sqrt3}-\frac{d_0^2}{3}\\
    &+\frac{2l_1d_0}{\sqrt3}+e^\gamma(q_0^2-p_0^2).
    \end{aligned}\label{eq:KKradialfactors}
\end{equation}
Applying the same substitution to the electric potential $A_t=-e^\gamma q_0(r+d_0/\sqrt3)e^{2\s}\mathcal P_e(0)/(\Delta\mathcal P_e)$. Consequently, the radial fields can be expressed as rational functions together with a square root involving the product of these two polynomials. The charged exterior with a positive factor requires
\begin{equation}
     \begin{gathered}
     0<\delta_e,\delta_m<1,\\
     1-2\delta_e+\delta_e\delta_m>0,\qquad
     1-2\delta_m+\delta_e\delta_m>0.
     \end{gathered}\label{eq:KKparameterdomain}
\end{equation}

In the Schwarzschild exterior, $e^{2\s}\in[0,1]$, with the horizon at $\s\to-\infty$ and spatial infinity at $\s=0$. The derivatives of $\mathcal P_e$ and $\mathcal P_m$ with respect to $e^{2\s}$ are $-2\delta_e(1-\delta_m e^{2\s})$ and $-2\delta_m(1-\delta_e e^{2\s})$. Both are negative in Eq. \eqref{eq:KKparameterdomain}. Hence the factors attain their positive minima at $\s=0$, which excludes a denominator zero throughout the exterior.  The horizon area equals $16\pi L^2/\sqrt{\mathcal P_e(0)\mathcal P_m(0)}$, the scalar there is $\sqrt3\,\ln[\mathcal P_e(0)/\mathcal P_m(0)]/4$, and the surface gravity is $\sqrt{\mathcal P_e(0)\mathcal P_m(0)}/(4L)$.

For the zero-energy harmonic function introduced above, applying the Coulomb substitution renders both factors positive everywhere in the exterior region. From their dominant terms and their ratio at the centre, one obtains
\[
     \operatorname{Area}_H=8\pi e^\gamma|q_0p_0|,\qquad
     \phi_H=\frac1{2\sqrt3}\ln\left|\frac{p_0}{q_0}\right|.
\]

A finite area therefore demands that both charges be present. Taking equal nonextremal polynomial parameters yields a balanced RN configuration, while setting one charge to zero produces the single-charge KK endpoint.

\paragraph{ModMax identification} \textbf{New explicit ModMax solution.} Its Maxwell metric coincides with the standard static KK dyon \cite{SJPoletti_1995,Rasheed:1995zv}. One can map it directly to Sec. IV of Bogush et al. \cite{Bogush:2020obx}—after converting their metric-signature convention—by taking
\[
    \begin{gathered}
    M=l_1,\qquad N=0,\qquad D=d_0/\sqrt3,\\
    Q=-e^{\gamma/2}q_0,\qquad P=-e^{\gamma/2}p_0.
    \end{gathered}
\]
The two polynomials in Eq. \eqref{eq:KKradialfactors} correspond to their metric functions $B$ and $A$, in that order, and our cubic relation matches their charge constraint. The scalar field agrees, their electric potential becomes ours after multiplication by $e^{\gamma/2}$, and their magnetic potential becomes ours after multiplication by $e^{-\gamma/2}$. Accordingly, this branch correspond to the ModMax extension.

\paragraph{Maxwell limit} For $\gamma=0$, the polynomial and radial formulas match the static KK dyon reconstructed in Ref. \cite{Bixano:2026xjg}. The dictionary to Bogush et al. \cite{Bogush:2020obx} given above determines the remaining sign choices and scalar conventions. The RN and one-charge endpoints are the previously identified limiting cases of this family, not distinct new dyons.

\section{A nonfrozen stationary family with a coupled scalar}\label{sec:active}

\subsection{Integration for an arbitrary harmonic}\label{sec:generalactive}

Now take complex $a(\s),b(\s)$ and real nonzero $c(\s)$, with $\al^2=1,3$. We impose
\[
    \begin{gathered}
    c=-\frac{\al d_0}{l_2}\Imm a,\qquad
    \Imm a=\frac{l_2}{L}f,\\
    \phi'=-\frac{d_0}{L}f,\qquad l_2d_0\ne0.
    \end{gathered}
\]
The second relation follows from $(\Imm a/f)'=0$. The first is an additional restriction relating the scalar and twist
along the harmonic trajectory. It selects the family studied here, it is not imposed on the other branches.

Differentiating $c=-\al d_0f/L$ and using $f'=2f\Ree a$ gives the scalar equation that this ansatz must satisfy:
\begin{equation}
    -\frac{2\al d_0}{L}f\Ree a
    =\frac{\al^2}{\epsilon_0}|b|^2
    \frac{\Ree(b^2)-\tanh\gamma|b|^2}
    {|b|^2-\tanh\gamma\Ree(b^2)}.
    \label{eq:activescalarequation}
\end{equation}
We retain this equation as a compatibility condition while integrating the metric.

The real gravitational equation and the energy give, respectively,
\[
    \begin{aligned}
    (\Ree a)'&=|b|^2-2(\Imm a)^2,\\
    |b|^2&=(\Ree a)^2+(\Imm a)^2
    +\frac{\epsilon_0d_0^2}{L^2}f^2-H.
    \end{aligned}
\]
Eliminating $|b|^2$ yields
\[
    (\Ree a)'=(\Ree a)^2
    +\frac{\epsilon_0d_0^2-l_2^2}{L^2}f^2-H.
\]
Since $(f^{-1/2})'=-\Ree a\,f^{-1/2}$, this becomes
\begin{equation}
    (f^{-1/2})''=Hf^{-1/2}
    +\frac{l_2^2-\epsilon_0d_0^2}{L^2}(f^{-1/2})^{-3}.
    \label{eq:activepinney}
\end{equation}
For $\mathcal D=f^{-1}$, differentiation cancels the inverse-power term and gives
\[
    \begin{gathered}
    \mathcal D'''=4H\mathcal D',\qquad
    \mathcal D(0)=1,\qquad \mathcal D'(0)=-2l_1/L,\\
    \mathcal D''(0)=2H+
    \frac{2(l_1^2+l_2^2-\epsilon_0d_0^2)}{L^2}.
    \end{gathered}
\]
The normalized solution is therefore
\[
    \begin{aligned}
    \mathcal D={}&\left[\cosh(h\s)-
    \frac{l_1}{hL}\sinh(h\s)\right]^2\\
    &+\frac{l_2^2-\epsilon_0d_0^2}{h^2L^2}\sinh^2(h\s),
    \qquad h^2=H,\\
    f={}&\mathcal D^{-1},\qquad
    a=-\frac{\mathcal D'}{2\mathcal D}
    +\frac{i l_2}{L\mathcal D},\qquad
    c=-\frac{\al d_0}{L\mathcal D}.
    \end{aligned}
\]
These expressions apply on a connected interval with $\mathcal D>0$.

Integrating $\phi'=-d_0/(L\mathcal D)$ with $\phi(0)=0$ gives three real forms. For $l_2^2>\epsilon_0d_0^2$,
\[
    \phi=-\frac{d_0}{\sqrt{l_2^2-\epsilon_0d_0^2}}
    \arctan\frac{\sqrt{l_2^2-\epsilon_0d_0^2}\sinh(h\s)}
    {hL\cosh(h\s)-l_1\sinh(h\s)}.
\]
For $l_2^2=\epsilon_0d_0^2$,
\[
    \phi=-\frac{d_0\sinh(h\s)}
    {hL\cosh(h\s)-l_1\sinh(h\s)}.
\]
For $l_2^2<\epsilon_0d_0^2$,
\[
    \begin{split}
    \phi={}&-\frac{d_0}{2\sqrt{\epsilon_0d_0^2-l_2^2}}\\
    &\times\ln\frac{hL\cosh(h\s)-
    [l_1-\sqrt{\epsilon_0d_0^2-l_2^2}]\sinh(h\s)}
    {hL\cosh(h\s)-
    [l_1+\sqrt{\epsilon_0d_0^2-l_2^2}]\sinh(h\s)}.
    \end{split}
\]
The arctangent is continued from $\s=0$ without jumps. Direct differentiation of each expression gives $-d_0/(L\mathcal D)$. The electromagnetic equations and the remaining scalar condition must still be imposed.

\subsubsection{Electromagnetic integration and charge compatibility}

For $\Imm a=l_2f/L$, Eq. \eqref{eq:twistcoupling} gives two first integrals. Their values at $\s=0$ fix
\[
    \psi'=\frac{2e^\gamma}{L\mathcal D\kappa^2}(q_0-l_2\chi),
    \qquad
    \chi'=\frac{2e^\gamma\kappa^2}{L\mathcal D}(p_0+l_2\psi).
\]
To integrate them, differentiate $\kappa\sqrt{\mathcal D}(2l_2A_t+p_0)$ twice and use Eq. \eqref{eq:activepinney}. The result is
\[
    \begin{split}
    &[\kappa\sqrt{\mathcal D}(2l_2A_t+p_0)]''\\
    &\quad=\left[H+
    \frac{(\al^2-\epsilon_0)d_0^2-(4e^{2\gamma}-1)l_2^2}
    {L^2\mathcal D^2}\right]\\
    &\qquad\times\kappa\sqrt{\mathcal D}(2l_2A_t+p_0).
    \end{split}
\]
The combination $\kappa^{-1}\sqrt{\mathcal D}(q_0-l_2\chi)$ satisfies the same equation. We now impose
\[
    (\al^2-\epsilon_0)d_0^2=(4e^{2\gamma}-1)l_2^2.
\]
This is an additional restriction that makes both equations linear with constant coefficient $H$. It selects an elementary subfamily and is not a necessary condition for every stationary solution of Eq. \eqref{eq:abc}.

At $\s=0$, the two combinations have values $p_0,q_0$ and derivatives
\[
    \begin{gathered}
    -\frac{p_0(l_1-\al d_0)-2e^\gamma l_2q_0}{L},\\
    -\frac{q_0(l_1+\al d_0)+2e^\gamma l_2p_0}{L}.
    \end{gathered}
\]
Solving the two linear equations with these data gives
\[
    \begin{aligned}
    A_t={}&\frac1{2l_2}\left\{
    \frac{e^{\al\phi}}{\sqrt{\mathcal D}}
    \left[p_0\cosh(h\s)\right.\right.\\
    &\left.\left.\hspace{4mm}-
    \frac{p_0(l_1-\al d_0)-2e^\gamma l_2q_0}{hL}
    \sinh(h\s)\right]-p_0\right\},\\
    \chi={}&\frac1{l_2}\left\{q_0-
    \frac{e^{-\al\phi}}{\sqrt{\mathcal D}}
    \left[q_0\cosh(h\s)\right.\right.\\
    &\left.\left.\hspace{4mm}-
    \frac{q_0(l_1+\al d_0)+2e^\gamma l_2p_0}{hL}
    \sinh(h\s)\right]\right\},\\
    \omega={}&\frac{2l_2}{L}\schdual,\qquad k=HK,\\
    A_\varphi={}&-\frac{2l_2A_t+p_0}{L}\schdual,
    \qquad\kappa=e^{-\al\phi}.
    \end{aligned}
\]
The original coefficient $b$ is then given by Eq. \eqref{eq:binverse}. Its phase can be read from
\[
    \begin{gathered}
    \frac{\Ree b}{\Imm b}=-e^\gamma\\[-1mm]
    \times\frac{q_0\cosh(h\s)-
    [q_0(l_1+\al d_0)+2e^\gamma l_2p_0]\frac{\sinh(h\s)}{hL}}
    {p_0\cosh(h\s)-
    [p_0(l_1-\al d_0)-2e^\gamma l_2q_0]\frac{\sinh(h\s)}{hL}}.
    \end{gathered}
\]

We must still check Eq. \eqref{eq:activescalarequation}. Substituting the first integrals reduces it to
\begin{equation}
    \begin{split}
    \epsilon_0Ld_0\mathcal D'={}&\al e^\gamma\mathcal D
    [\kappa^{-2}(q_0-l_2\chi)^2\\
    &\hspace{12mm}-\kappa^2(2l_2A_t+p_0)^2].
    \end{split}\label{eq:activescalaridentity}
\end{equation}
Evaluation at $\s=0$ gives the condition below. Differentiating once at $\s=0$ and using the energy in Eq. \eqref{eq:charges} gives the charge-product condition. Together with the linearization condition, they are
\begin{equation}
    \begin{aligned}
    (\al^2-\epsilon_0)d_0^2
    &=(4e^{2\gamma}-1)l_2^2,\\
    \al e^\gamma(q_0^2-p_0^2)
    &=-2\epsilon_0l_1d_0,\\
    4\al e^{2\gamma}l_2q_0p_0
    &=-d_0[2\epsilon_0l_2^2+
    (\al^2-\epsilon_0)e^\gamma(q_0^2+p_0^2)].
    \end{aligned}\label{eq:activecompatibility}
\end{equation}
These conditions are also sufficient for the scalar identity. Indeed, the third derivative of each side of Eq. \eqref{eq:activescalaridentity} equals $4H$ times its first derivative: the left side follows from $\mathcal D'''=4H\mathcal D'$, and the right side is a difference of squares of solutions of the two linear electromagnetic equations. Their values and first two derivatives at $\s=0$ agree under Eq. \eqref{eq:activecompatibility}. Uniqueness therefore proves the identity throughout the interval.

Finally, the energy requires
\[
    |b|^2=\frac{\mathcal D'^2}{4\mathcal D^2}
    +\frac{l_2^2+\epsilon_0d_0^2}{L^2\mathcal D^2}-H.
\]
Using Eqs. \eqref{eq:bnorm} and \eqref{eq:activescalaridentity}, the difference between this expression and the norm reconstructed from the potentials has derivative $2\Ree a$ times itself. It vanishes at $\s=0$ by Eq. \eqref{eq:charges}, and hence vanishes everywhere. Thus the gravitational, electromagnetic and scalar equations are all satisfied.

For $H=0$, the arbitrary harmonic denominator is 
\[
    \mathcal D=1-\frac{2l_1}{L}\s
    +\frac{l_1^2+l_2^2-\epsilon_0d_0^2}{L^2}\s^2.
\]
The compatibility conditions remain unchanged.

\subsection{The Schwarzschild harmonic and the black-hole domain}

With $H=h^2>0$, the denominator behaves as $f\propto e^{2h\s}$ near to the horizon, while $\phi'\to0$. Imposing the area condition sets $h=1$. Using the Schwarzschild substitution yields $\mathcal D=(r^2+l_2^2-\epsilon_0d_0^2)/\Delta$, and therefore
\begin{equation}
     \begin{aligned}
     f={}&\frac{(r-l_1)^2-L^2}{r^2+l_2^2-\epsilon_0d_0^2},\\
     \dd s^2={}&-f(\dd t-2l_2\cos\theta\dd\varphi)^2
     +f^{-1}\dd r^2\\
     &+(r^2+l_2^2-\epsilon_0d_0^2)\dd\Omega_2^2,\\
     A_t={}&\frac1{2l_2}\left[
     e^{\al\phi}\frac{p_0(r-\al d_0)-2e^\gamma l_2q_0}
     {\sqrt{r^2+l_2^2-\epsilon_0d_0^2}}-p_0\right],\\
     A_\varphi={}&-(2l_2A_t+p_0)\cos\theta,\\
     \omega={}&2l_2\cos\theta,
     \qquad k=K[\s].
     \end{aligned}\label{eq:activeBHfields}
\end{equation}

The scalar field is
\begin{equation}
     \phi=\frac{d_0}{\sqrt{l_2^2-\epsilon_0d_0^2}}
     \arctan\frac{\sqrt{l_2^2-\epsilon_0d_0^2}}{r}.
     \label{eq:activeBHscalar}
\end{equation}

When $l_2^2=\epsilon_0d_0^2$, the corresponding real limiting value is $d_0/r$. If instead $l_2^2<\epsilon_0d_0^2$, it becomes $d_0\ln\!\left[(r+\sqrt{\epsilon_0d_0^2-l_2^2})/(r-\sqrt{\epsilon_0d_0^2-l_2^2})\right]/\left[2\sqrt{\epsilon_0d_0^2-l_2^2}\right]$. Moreover, every derivative equals $-d_0/(r^2+l_2^2-\epsilon_0d_0^2)$.

For $l_1\ge0$, a regular exterior requires $L>0$ and
\[
     r_+^2+l_2^2-\epsilon_0d_0^2>0.
\]

As the radial denominator grows, it consequently rules out all external scalar poles. Removing the charge sum, difference, and product from Eq. \eqref{eq:activecompatibility} as well as from the energy expression yields
\begin{equation}
     \begin{split}
     &[\al^2+\epsilon_0(4e^{2\gamma}-1)]
     (l_1^2+l_2^2-\epsilon_0d_0^2-HL^2)^2\\
     &\hspace{20mm}=16e^{2\gamma}d_0^2HL^2.
     \end{split}\label{eq:activeenergyidentity}
\end{equation}

At $H=1$, real charges and a positive horizon denominator gives
\begin{equation}
     \begin{split}
     L={}&\sqrt{l_1^2+
     \frac{\al^2l_2^2}{\al^2+\epsilon_0(4e^{2\gamma}-1)}}\\
     &-\frac{2e^\gamma|d_0|}
     {\sqrt{\al^2+\epsilon_0(4e^{2\gamma}-1)}}>0.
     \end{split}\label{eq:activeBHroot}
\end{equation}

The expression under the square root must be positive. Choosing the alternative algebraic sign leads either to a negative squared charge or to a nonpositive denominator in the horizon term. For the root we have chosen,
\[
     \begin{aligned}
     &r_+^2+l_2^2-\epsilon_0d_0^2\\
     &\quad=2L\left[l_1+
     \sqrt{l_1^2+\frac{\al^2l_2^2}{\al^2+\epsilon_0(4e^{2\gamma}-1)}}\right].
     \end{aligned}
\]

Direct evaluation therefore gives
\[
     \begin{aligned}
     \operatorname{Area}_+&=4\pi(r_+^2+l_2^2-\epsilon_0d_0^2),\\
     \varkappa_+&=\frac{L}{r_+^2+l_2^2-\epsilon_0d_0^2},
     \qquad T_+=\frac{\varkappa_+}{2\pi}.
     \end{aligned}
\]

For a positive denominator, the fields remain analytic. The ingoing radial form of the metric is $-f(\dd v-2l_2\cos\theta\dd\varphi)^2+2\dd r(\dd v-2l_2\cos\theta\dd\varphi)$, it stays regular when $f$ has a simple zero, thereby establishing a local Killing horizon.

For the Schwarzschild harmonic, $\omega=2l_2\cos\theta$ identifies $l_2$ as a NUT parameter. This representation has no independent Kerr rotation parameter. Its asymptotic expansion gives $f=1-2l_1/r+O(r^{-2})$ and $\phi=d_0/r+O(r^{-2})$.
Thus $l_1$ and $d_0$ are the mass and scalar parameters in the chosen normalization. For $l_1\ne0$, the second compatibility condition fixes
\[
    d_0=-\frac{\al e^\gamma(q_0^2-p_0^2)}{2\epsilon_0l_1}.
\]
The scalar charge is therefore constrained by the mass and electromagnetic charges, rather than being independent hair. Its invariant gradient is
\[
    (\nabla\phi)^2=
    \frac{\Delta d_0^2}{(r^2+l_2^2-\epsilon_0d_0^2)^3},
\]
which tends to zero at the horizon when its denominator is positive. The scalar is nonconstant outside while remaining finite on the horizon. On the two axes, $\omega(0)-\omega(\pi)=4l_2\ne0$, so a single time gauge cannot make $\omega$ vanish on both. The regular horizon established above is local and does not remove this standard global NUT condition.

With standard unit coupling, the first charge constraint implies $\gamma=-\ln2$. The other constraints read
\[
     \begin{aligned}
     q_0^2-p_0^2&=-4l_1d_0,\qquad q_0p_0=-2l_2d_0,\\
     q_0^2+p_0^2&=4|d_0|\sqrt{l_1^2+l_2^2},\\
     L&=\sqrt{l_1^2+l_2^2}-|d_0|,
     \qquad 0<|d_0|<\sqrt{l_1^2+l_2^2}.
     \end{aligned}
\]

The area is given by $8\pi L[l_1+\sqrt{l_1^2+l_2^2}]$, while the temperature is $1/[4\pi(l_1+\sqrt{l_1^2+l_2^2})]$. Holding the mass and NUT charge fixed, introducing the scalar reduces the area but leaves the temperature unchanged. At the upper endpoint of scalar charge, the area goes to zero and the scalar field diverges.

For ordinary KK coupling the nonextremal domain is
\[
     d_0^2=\frac{4e^{2\gamma}-1}{2}l_2^2,\qquad
     \frac14<e^{2\gamma}<\frac34+\frac{l_1^2}{2l_2^2}.
\]

The strict lower bound ensures $d_0^2>0$, equality would remove the scalar. The strict upper bound ensures $L>0$. For the usual physical ModMax convention, intersect this range with $\gamma\le0$. When $e^{2\gamma}=3/4$, the scalar transitions smoothly among its arctangent, rational, and logarithmic expressions on a regular exterior.

For the phantom sign the domain is
\[
     \frac14<e^{2\gamma}<\frac{\al^2+1}{4},\qquad
     d_0^2=\frac{4e^{2\gamma}-1}{\al^2+1}l_2^2.
\]

The denominator remains positive throughout. At the upper endpoint, where $H=0$, the Coulomb harmonic yields
\[
     \begin{aligned}
     f&=\frac{(r-l_1)^2}{r^2+l_2^2+d_0^2},\qquad
     d_0^2=\frac{\al^2}{\al^2+1}l_2^2,\\
     \operatorname{Area}_+&=4\pi(l_1^2+l_2^2+d_0^2),
     \qquad T_+=0.
     \end{aligned}
\]

The limiting scalar and four-potential in Eqs. \eqref{eq:activeBHfields}--\eqref{eq:activeBHscalar} have a strictly positive denominator. For the degenerate horizon, $\gamma=-\tfrac12\ln2$ at unit coupling, whereas $\gamma=0$ at KK coupling.

For the standard $H=0$ boundary one finds $d_0^2=l_1^2+l_2^2$ and $\mathcal D=1-2l_1\s/L$. The scalar field blows up at the Coulomb centre and the curvature becomes singular. Hence, it is not regarded as a regular extremal black hole.

\paragraph{ModMax identification}
\textbf{New nonfrozen ModMax family.} This defines a restricted branch for $\gamma\ne0$, $l_2d_0\ne0$, within the stated horizon domain. The electromagnetic numerators have a nonzero Wronskian exactly when $e^\gamma l_2(q_0^2+p_0^2)+\al d_0q_0p_0\ne0$. The charge conditions then yield
\[
     \begin{split}
     &e^\gamma l_2(q_0^2+p_0^2)+\al d_0q_0p_0=\frac{l_2}{4e^{2\gamma}}
     (l_1^2+l_2^2-\epsilon_0d_0^2-HL^2).
     \end{split}
\]

Its right-hand side cannot be zero on the nonextremal region, since Eq. \eqref{eq:activeenergyidentity} would then force $d_0^2L^2=0$. Therefore, the phase is not constant and must vary. These fields are not the same as the frozen-invariant solution described in Ref. \cite{Bixano:2026uqk}. 
This distinction follows from the varying electromagnetic phase proved above, not from the scalar being nonconstant alone.
In Ref. \cite{Ayon-Beato:2024vph}, the scalar $\Phi$ instead obeys $\Box\Phi-R\Phi/6=4\lambda\Phi^3$ and is conformally coupled to curvature. Here the scalar source comes from $e^{-2\al\phi}\mathcal L_{\rm MM}$ in Eq. \eqref{eq:action}. They are therefore solutions of different scalar theories. The Maxwell comparison for the same dilatonic coupling is given below.

\paragraph{Maxwell limit}
For the ordinary scalar, $\epsilon_0=+1$, at $\al^2=1$, setting $\gamma=0$ in the first compatibility condition gives $0=3l_2^2$. Thus this nonzero-NUT family has no Maxwell limit within the stated ansatz. This conclusion does not apply to the phantom sign or to stationary families outside this ansatz.

\textbf{Known Maxwell limit at ordinary KK coupling.} At $\gamma=0$,
\begin{equation}
     d_0^2=\frac32l_2^2,\qquad
     L=\sqrt{l_1^2+\frac12l_2^2}-|l_2|,
     \qquad l_1>\frac{|l_2|}{\sqrt2}.
     \label{eq:activeKKMaxwell}
\end{equation}

This is the restricted NUT dyon of Bogush et al., \cite{Bogush:2020obx}, with
\[
     \begin{gathered}
     M=l_1,\quad N=-l_2,\quad D=d_0/\sqrt3,\\
     Q=-q_0,\qquad P=-p_0.
     \end{gathered}
\]

This yields $N^2=2D^2$ and $Q^2-P^2=-2MD$. The two metric polynomials simplify to $(r+D)^2$ and $(r-D)^2$, and the scalar field matches Eq. \eqref{eq:activeBHscalar}. The electromagnetic field then becomes
\[
     A_t=-\frac{q_0(r+D)+l_2p_0}{(r-D)^2},
\]
with an identical angular potential. The mass term in Eq. \eqref{eq:activeKKMaxwell} maintains the nonextremal horizon. By contrast, the phantom case that has $d_0^2=3l_2^2/4$ and lies in the phantom Maxwell sector.


\clearpage
\onecolumngrid



\begin{turnpage}
\setlength{\parindent}{0pt}
\begin{table}[p]
\centering
\caption{Charged harmonic ModMax fields and their black-hole domains. Coupling cases and changes of harmonic or LWP frame are not counted as independent branches. NUT horizons are local; the axial exterior retains its constitutive-domain qualification.}\label{tab:catalogue}
\small
\setlength{\tabcolsep}{3pt}
\begin{tabular}{llllll}
\hline
\parbox[t]{1.8cm}{\raggedright Branch / representation}
&
\parbox[t]{2.8cm}{\raggedright Horizon domain}
&
\parbox[t]{6.5cm}{\raggedright Functions for arbitrary $\mathfrak s$}
&
\parbox[t]{2.5cm}{\raggedright Identification}
&
\parbox[t]{2.7cm}{\raggedright Maxwell limit}
&
\parbox[t]{5.6cm}{\raggedright Schwarzschild-harmonic representation}\\
\hline
\parbox[t]{1.8cm}{\raggedright RN sector\newline Sec. \ref{sec:RN}}
&
\parbox[t]{2.8cm}{\raggedright $\al=0$, $\phi=l_2=0$.\newline $\mathcal D>0$, $H=h^2$.\newline A nonextremal horizon requires $h=1$, $l_1>0$, $L^2=l_1^2-e^\gamma(q_0^2+p_0^2)>0$.}
&
\parbox[t]{6.5cm}{\raggedright $\mathcal D=\cosh(h\s)-l_1\sinh(h\s)/(hL)$.\newline
$f=\mathcal D^{-2}$, $\omega=0$, $k=HK$.\newline
$A_t=e^\gamma q_0\sinh(h\s)/(hL\mathcal D)$,\newline
$A_\varphi=-p_0\schdual/L$, $\phi=0$.\newline The $h=0$ limit is linear.}
&
\parbox[t]{2.5cm}{\raggedright \textbf{Known solution}}
&
\parbox[t]{2.7cm}{\raggedright Known dyonic RN.\newline Set $\gamma=0$ in the charge square and electric potential.}
&
\parbox[t]{5.6cm}{\raggedright $\mathcal D=r/\sqrt\Delta$.\newline $f=\Delta/r^2$, $\omega=0$, $k=K$, $A_t=-e^\gamma q_0/r$,\newline $A_\varphi=-p_0\cos\theta$, $\phi=0$.\newline $r_+=l_1+L$. The degenerate limit uses the Coulomb harmonic.}\\
\noalign{\smallskip}
\parbox[t]{1.8cm}{\raggedright Complex branch\newline Electric LWP\newline Sec. \ref{sec:NUT}}
&
\parbox[t]{2.8cm}{\raggedright $\al=0$, $\phi=0$, $l_2\ne0$.\newline $\mathcal D$ is complex.\newline The local nonextremal horizon has $h=1$ and $L^2=l_1^2+l_2^2-e^\gamma(q_0^2+p_0^2)>0$.}
&
\parbox[t]{6.5cm}{\raggedright $\mathcal D=\cosh(h\s)-(l_1+il_2)\sinh(h\s)/(hL)$.\newline
$f=|\mathcal D|^{-2}$, $\omega=2l_2\schdual/L$, $k=HK$.\newline
$A_t=\Imm\{(q_0+ip_0)[(\bar{\mathcal D}/\mathcal D)^{e^\gamma}-1]\}/(2l_2)$.\newline
$A_\varphi=-\omega A_t-p_0\schdual/L$, $\phi=0$.}
&
\parbox[t]{2.5cm}{\raggedright \textbf{Known solution}}
&
\parbox[t]{2.7cm}{\raggedright Known Maxwell charged NUT.\newline The phase exponent is one. The degenerate member also has a one-harmonic IWP identification.}
&
\parbox[t]{5.6cm}{\raggedright $f=\Delta/(r^2+l_2^2)$, $\omega=2l_2\cos\theta$, $k=K$.\newline In $A_t$, use $\bar{\mathcal D}/\mathcal D=(r-il_2)/(r+il_2)$.\newline $A_\varphi=-(2l_2A_t+p_0)\cos\theta$, $\phi=0$.\newline Local horizon, NUT axis treated separately.}\\
\hline
\end{tabular}
\end{table}
\end{turnpage}

\begin{turnpage}
\setlength{\parindent}{0pt}
\begin{table}[p]
\centering
\caption{Charged harmonic ModMax fields and their black-hole domains (continued).}
\small
\setlength{\tabcolsep}{3pt}
\begin{tabular}{llllll}
\hline
\parbox[t]{1.8cm}{\raggedright Branch / representation}
&
\parbox[t]{2.8cm}{\raggedright Horizon domain}
&
\parbox[t]{6.5cm}{\raggedright Functions for arbitrary $\mathfrak s$}
&
\parbox[t]{2.5cm}{\raggedright Identification}
&
\parbox[t]{2.7cm}{\raggedright Maxwell limit}
&
\parbox[t]{5.6cm}{\raggedright Schwarzschild-harmonic representation}\\
\hline
\parbox[t]{1.8cm}{\raggedright Single charge\newline Generic coupling\newline Sec. \ref{sec:GM}}
&
\parbox[t]{2.8cm}{\raggedright $\al^2=1,3$ with $1+\epsilon_0\al^2\ne0$.\newline $\lambda=1$, $d_0=\epsilon_0\al(L-l_1)$, and $r_+-\al d_0>0$.\newline Domains in Eqs. \eqref{eq:GMregularity}, \eqref{eq:GMphantomBH}.}
&
\parbox[t]{6.5cm}{\raggedright $f=\exp[2\al(d_0+\epsilon_0\al l_1)\s/((1+\epsilon_0\al^2)L)]\mathcal D_e^{-2/(1+\epsilon_0\al^2)}$.\newline
$\omega=0$, $k=HK$, $A_\varphi=0$.\newline
$A_t=e^\gamma q_0\sinh(\lambda\s)/(\lambda L\mathcal D_e)$.\newline
$\phi=-\epsilon_0\al\ln\mathcal D_e/(1+\epsilon_0\al^2)-(d_0+\epsilon_0\al l_1)\s/((1+\epsilon_0\al^2)L)$.}
&
\parbox[t]{2.5cm}{\raggedright \textbf{Known solution}}
&
\parbox[t]{2.7cm}{\raggedright Known Gibbons--Maeda/GHS or phantom EMD profile.\newline Dictionary: Eq. \eqref{eq:GMdictionary}.}
&
\parbox[t]{5.6cm}{\raggedright $f$ and $\phi$ are Eq. \eqref{eq:GMradial}.\newline $\omega=0$, $k=K$, $A_t=-e^\gamma q_0/(r-\al d_0)$, $A_\varphi=0$.\newline $r_+=l_1+L$ and $r_+-\al d_0>0$.\newline Ordinary charged extremal area: zero.}\\
\noalign{\smallskip}
\parbox[t]{1.8cm}{\raggedright Single charge\newline Exceptional case\newline Sec. \ref{sec:phantomunit}}
&
\parbox[t]{2.8cm}{\raggedright $\epsilon_0=-1$, $\al=1$.\newline Horizon: $\lambda=1$, $d_0=l_1-L$, $e^\gamma q_0^2=2L(l_1-L)$, $0<L<l_1$.}
&
\parbox[t]{6.5cm}{\raggedright $f=\exp[2l_1\s/L+e^\gamma q_0^2(e^{2\lambda\s}-1-2\lambda\s)/(2\lambda^2L^2)]$.\newline
$\omega=0$, $k=HK$.\newline $A_t=e^\gamma q_0(e^{2\lambda\s}-1)/(2\lambda L)$, $A_\varphi=0$, $\phi=\lambda\s-\tfrac12\ln f$.}
&
\parbox[t]{2.5cm}{\raggedright \textbf{New ModMax family}}
&
\parbox[t]{2.7cm}{\raggedright Known exceptional phantom black hole of Cl\'ement--Fabris--Rodrigues \cite{Clement:2009ai}.}
&
\parbox[t]{5.6cm}{\raggedright $f=\frac{r-l_1-L}{r-l_1+L}e^{-2d_0/(r-l_1+L)}$, $\omega=0$, $k=K$.\newline $A_t=-e^\gamma q_0/(r-l_1+L)$, $A_\varphi=0$, $\phi=d_0/(r-l_1+L)$.\newline The denominator is at least $2L>0$.}\\
\hline
\end{tabular}
\end{table}
\end{turnpage}

\begin{turnpage}
\setlength{\parindent}{0pt}
\begin{table}[p]
\centering
\caption{Charged harmonic ModMax fields and their black-hole domains (continued).}
\small
\setlength{\tabcolsep}{3pt}
\begin{tabular}{llllll}
\hline
\parbox[t]{1.8cm}{\raggedright Branch / representation}
&
\parbox[t]{2.8cm}{\raggedright Horizon domain}
&
\parbox[t]{6.5cm}{\raggedright Functions for arbitrary $\mathfrak s$}
&
\parbox[t]{2.5cm}{\raggedright Identification}
&
\parbox[t]{2.7cm}{\raggedright Maxwell limit}
&
\parbox[t]{5.6cm}{\raggedright Schwarzschild-harmonic representation}\\
\hline
\parbox[t]{1.8cm}{\raggedright Unit dyon\newline Sec. \ref{sec:string}}
&
\parbox[t]{2.8cm}{\raggedright $\epsilon_0=+1$, $\al=1$.\newline $h_e=h_m=1$ and Eq. \eqref{eq:stringcharges}.\newline $l_1>e^{\gamma/2}(|q_0|+|p_0|)/\sqrt2$.}
&
\parbox[t]{6.5cm}{\raggedright $f=(\mathcal D_e\mathcal D_m)^{-1}$, $\omega=0$, $k=HK$.\newline
$A_t=e^\gamma q_0\sinh(h_e\s)/(h_eL\mathcal D_e)$, $A_\varphi=-p_0\schdual/L$.\newline
$\phi=\tfrac12\ln(\mathcal D_m/\mathcal D_e)$.\newline
Denominators in Eq. \eqref{eq:stringdenominators}, $H=(h_e^2+h_m^2)/2$.}
&
\parbox[t]{2.5cm}{\raggedright \textbf{New ModMax family}}
&
\parbox[t]{2.7cm}{\raggedright Known unit-coupling EMD dyon \cite{Gibbons:1987ps,SJPoletti_1995}.\newline The mapping is Eqs. \eqref{eq:staticembeddingcharges}--\eqref{eq:staticembedding}.}
&
\parbox[t]{5.6cm}{\raggedright $f=\Delta/(r^2-d_0^2)$, $\omega=0$, $k=K$.\newline $A_t=-e^\gamma q_0/(r-d_0)$, $A_\varphi=-p_0\cos\theta$, $\phi=\tfrac12\ln[(r+d_0)/(r-d_0)]$.\newline $r_+>|d_0|$. The two-charge degenerate area is $8\pi e^\gamma|q_0p_0|$.}\\
\noalign{\smallskip}
\parbox[t]{1.8cm}{\raggedright KK dyon\newline Sec. \ref{sec:KK}}
&
\parbox[t]{2.8cm}{\raggedright $\epsilon_0=+1$, $\al^2=3$, $h=1$.\newline Positive-factor domain in Eq. \eqref{eq:KKparameterdomain}, cubic scalar constraint in Eq. \eqref{eq:KKconstraint}.}
&
\parbox[t]{6.5cm}{\raggedright $f=e^{2h\s}\sqrt{\mathcal P_e(0)\mathcal P_m(0)/(\mathcal P_e\mathcal P_m)}$, $\omega=0$, $k=h^2K$.\newline
$A_t=\frac{hL}{q_0}[(1-\delta_e e^{2h\s})/\mathcal P_e-(1-\delta_e)/\mathcal P_e(0)]$.\newline
$A_\varphi=-p_0\schdual/L$, $\phi=-\frac{\sqrt3}{4}\ln[\mathcal P_e\mathcal P_m(0)/(\mathcal P_m\mathcal P_e(0))]$.}
&
\parbox[t]{2.5cm}{\raggedright \textbf{New ModMax family}}
&
\parbox[t]{2.7cm}{\raggedright Known static KK dyon \cite{SJPoletti_1995,Rasheed:1995zv}.\newline Same polynomial parameters, exact charge and potential mapping in Eqs. \eqref{eq:staticembeddingcharges}--\eqref{eq:staticembedding}.}
&
\parbox[t]{5.6cm}{\raggedright Set $e^{2\s}=(r-l_1-L)/(r-l_1+L)$ in every polynomial and physical function.\newline $\omega=0$, $k=K$, $A_\varphi=-p_0\cos\theta$.\newline The two-charge extremal fields use Eqs. \eqref{eq:KKextremefactors}--\eqref{eq:KKextremeAt}, their area is $8\pi e^\gamma|q_0p_0|$.}\\
\hline
\end{tabular}
\end{table}
\end{turnpage}

\begin{turnpage}
\setlength{\parindent}{0pt}
\begin{table}[p]
\centering
\caption{Charged harmonic ModMax fields and their black-hole domains (continued).}
\small
\setlength{\tabcolsep}{3pt}
\begin{tabular}{llllll}
\hline
\parbox[t]{1.8cm}{\raggedright Branch / representation}
&
\parbox[t]{2.8cm}{\raggedright Horizon domain}
&
\parbox[t]{6.5cm}{\raggedright Functions for arbitrary $\mathfrak s$}
&
\parbox[t]{2.5cm}{\raggedright Identification}
&
\parbox[t]{2.7cm}{\raggedright Maxwell limit}
&
\parbox[t]{5.6cm}{\raggedright Schwarzschild-harmonic representation}\\
\hline
\parbox[t]{1.8cm}{\raggedright Active stationary scalar\newline Sec. \ref{sec:active}}
&
\parbox[t]{2.8cm}{\raggedright $l_2d_0\ne0$, $\al^2=1,3$, either sign.\newline Three compatibilities in Eq. \eqref{eq:activecompatibility}.\newline Horizon root in Eq. \eqref{eq:activeBHroot} and the stated coupling domains.}
&
\parbox[t]{6.5cm}{\raggedright $f=\mathcal D^{-1}$, $\omega=2l_2\schdual/L$, $k=HK$.\newline
$A_t=\{e^{\al\phi}[p_0\cosh(h\s)-[p_0(l_1-\al d_0)-2e^\gamma l_2q_0]\sinh(h\s)/(hL)]/\sqrt{\mathcal D}-p_0\}/(2l_2)$.\newline
$A_\varphi=-(2l_2A_t+p_0)\schdual/L$.\newline
$\phi=-\frac{d_0}{\sqrt{l_2^2-\epsilon_0d_0^2}}\arctan\frac{\sqrt{l_2^2-\epsilon_0d_0^2}\sinh(h\s)}{hL\cosh(h\s)-l_1\sinh(h\s)}$.\newline Real rational and logarithmic forms are given in the text.}
&
\parbox[t]{2.5cm}{\raggedright \textbf{New ModMax family}}
&
\parbox[t]{2.7cm}{\raggedright Ordinary $\al^2=1$: no nonzero-NUT Maxwell limit.\newline Ordinary $\al^2=3$: known NUT KK subfamily \cite{Bogush:2020obx}.\newline Phantom KK: the stated degenerate Maxwell endpoint.}
&
\parbox[t]{5.6cm}{\raggedright $f=\Delta/(r^2+l_2^2-\epsilon_0d_0^2)$, $\omega=2l_2\cos\theta$, $k=K$.\newline
$A_t=\{e^{\al\phi}[p_0(r-\al d_0)-2e^\gamma l_2q_0]/\sqrt{r^2+l_2^2-\epsilon_0d_0^2}-p_0\}/(2l_2)$.\newline
$A_\varphi=-(2l_2A_t+p_0)\cos\theta$, $\phi=\frac{d_0}{\sqrt{l_2^2-\epsilon_0d_0^2}}\arctan\frac{\sqrt{l_2^2-\epsilon_0d_0^2}}r$.\newline Use its real limits when required. The horizon is locally regular.}\\
\noalign{\smallskip}
\parbox[t]{1.8cm}{\raggedright \label{tab:catalogue:end}Complex branch\newline Magnetic LWP\newline Sec. \ref{sec:swirlBH}}
&
\parbox[t]{2.8cm}{\raggedright $\al=0$, $\phi=0$, $l_1=L>0$.\newline $E_{\rm B}^2+B_{\rm B}^2>0$.\newline $j\gamma\ne0$ gives the new exterior.\newline Smooth horizon limit, constitutive continuation at its null circle remains open.}
&
\parbox[t]{6.5cm}{\raggedright $f_{\rm m}=L^2e^{2\s}/F_{\rm B}$, $\omega_{\rm m}=-4j\schdual$, $e^{2k_{\rm m}}=L^2e^{2K}$.\newline
$f=\rho^2/f_{\rm m}-f_{\rm m}\omega_{\rm m}^2$, $\omega=-f_{\rm m}\omega_{\rm m}/f$, $e^{2k}=fe^{2k_{\rm m}}/f_{\rm m}$.\newline
$A_\varphi$ is the elementary sine/cosine in Eq. \eqref{eq:axialAphi},\newline
$A_t=(e^\gamma E_{\rm B}+4jA_\varphi)\schdual$, $\phi=0$.}
&
\parbox[t]{2.5cm}{\raggedright \textbf{New ModMax family}}
&
\parbox[t]{2.7cm}{\raggedright Known Schwarzschild in the electromagnetic swirling universe \cite{Barrientos:2024pkt}.\newline $M=l_1$, $E_{\rm ref}=-E_{\rm B}$, $B_{\rm ref}=-B_{\rm B}$, $j_{\rm ref}=j$.}
&
\parbox[t]{5.6cm}{\raggedright Insert the axial Schwarzschild harmonic in Eq. \eqref{eq:swirlBHharmonic}:\newline
$L^2e^{2\s}=r^2\sin^2\theta$, $\schdual=(r-2l_1)\cos\theta$.\newline
Use Eq. \eqref{eq:axialmetric} with this harmonic, $A_t=(e^\gamma E_{\rm B}+4jA_\varphi)(r-2l_1)\cos\theta$.\newline
$r_+=2l_1$, $\operatorname{Area}_+=16\pi l_1^2$, $\varkappa_+=1/(4l_1)$.}\\
\hline
\end{tabular}
\end{table}
\end{turnpage}

\clearpage
\twocolumngrid

\section{Conclusions}

The catalogue has eight entries: three known ModMax solutions and five new configurations. The known ones are the RN family, the charged Taub--NUT family, and the generic single-charge construction. The new ones are the exceptional unit-coupling phantom solution, static dyons at unit and KK couplings, a stationary family with an active scalar, and the axial Schwarzschild exterior in a Melvin--swirling background. Extremal limits and alternative harmonic or LWP forms stay within these branches and are not counted as new families.

We proved two theorems. The first gives a necessary and sufficient condition for a constant electromagnetic phase along the harmonic trajectory and shows that a static dyon with an active scalar may still have a varying phase. The second establishes an exact static correspondence with Maxwell–scalar fields, including the scalar source and the distinct rescalings of the electric and angular potentials, and shows that charged twisting solutions cannot be generated by constant diagonal rescalings of the electromagnetic potentials with the metric, scalar, and twist fixed. Thus static ModMax families keep the known Maxwell metric and scalar profiles, while stationary solutions require solving the coupled ModMax equations.

The reconstructed fields also determine the black-hole domains. For the nonextremal Schwarzschild harmonic, finite nonzero horizon area requires $H=1$, together with finite matter fields and positive denominators throughout the exterior.
For the ordinary scalar, the stationary unit-coupling family requires $\gamma=-\ln2$ and has no nonzero-NUT Maxwell limit within the stated ansatz. At KK coupling, its Maxwell limit is a known restricted NUT dyon. The scalar charge is constrained by the compatibility conditions rather than being independent hair, and the NUT horizon is locally regular
in the stated domain. The axial construction gives an exact ModMax exterior with finite metric and electromagnetic limits at the horizon. Its constitutive continuation through the null equatorial circle remains a separate problem.

\begin{acknowledgments}
We thank Carlos A. R. Herdeiro for his helpful comments and discussions on the solutions, their Maxwell limits and their identification. L.B. also thanks him and the University of Aveiro, Portugal, for their hospitality during his research visit, where these discussions took place.
\end{acknowledgments}

\bibliographystyle{elsarticle-harv} 
\bibliography{Bibliografia}

\end{document}